\documentclass[fleqn,usenatbib]{mnras}

\usepackage{newtxtext,newtxmath}
\usepackage[T1]{fontenc}

\DeclareRobustCommand{\VAN}[3]{#2}
\let\VANthebibliography\thebibliography
\def\thebibliography{\DeclareRobustCommand{\VAN}[3]{##3}\VANthebibliography}

\usepackage{graphicx}	% Including figure files
\usepackage{amsmath}	% Advanced maths commands
\usepackage{amssymb}	% Extra maths symbols
\usepackage{subcaption}
\newcommand{\nus}{{\em NuSTAR}}
\newcommand{\mwd}{$M_{\rm WD}$}
\newcommand{\msun}{$M_{\odot}$}
\newcommand{\rwd}{$R_{\rm WD}$}
\newcommand{\rmag}{$R_{\rm m}$}
\newcommand{\rco}{$R_{\rm co}$}
\newcommand{\pspin}{$P_{\rm spin}$}
\newcommand{\mdot}{$\dot{M}$}

\title[Magnetic white dwarf masses with NuSTAR]{Measuring the masses of magnetic white dwarfs: A {\em NuSTAR} Legacy Survey}

\author[A. W. Shaw et al.]{
A. W. Shaw,$^{1}$\thanks{E-mail: aarrans@unr.edu}
C. O. Heinke,${^2}$
K. Mukai,$^{3,4}$
J. A. Tomsick,$^{5}$
V. Doroshenko,$^{6,7}$
\newauthor V. F. Suleimanov,$^{6,8,7}$
D. J. K. Buisson,$^{9}$
P. Gandhi,$^{9}$
B. W. Grefenstette,$^{10}$
J. Hare,$^{11,12}$
\newauthor J. Jiang,$^{13,14}$
R. M. Ludlam,$^{10}$
V. Rana,$^{15}$
G. R. Sivakoff$^{2}$
\\
$^{1}$Department of Physics, University of Nevada, Reno, NV 89557, USA\\
$^{2}$Dept.\ of Physics, University of Alberta, CCIS 4-183, Edmonton, AB T6G 2E1, Canada\\
$^{3}$CRESST and X-ray Astrophysics Laboratory, NASA Goddard Space Flight Center, Greenbelt, MD 20771, USA\\
$^{4}$Department of Physics, University of Maryland, Baltimore County, 1000 Hilltop Circle, Baltimore, MD 21250, USA\\
$^{5}$Space Sciences Laboratory, 7 Gauss Way, University of California, Berkeley, CA 94720-7450, USA\\
$^{6}$Institut f\"{u}r Astronomie und Astrophysik, Kepler Center for Astro and Particle Physics, Eberhard Karls Universit\"{a}t, Sand 1, 72076 T\"{u}bingen, Germany\\
$^{7}$Space Research Institute of the Russian Academy of Sciences, Profsoyuznaya Str. 84/32, Moscow 117997, Russia\\
$^{8}$Kazan (Volga region) Federal University, Kremlevskaya str. 18, Kazan 420008, Russia\\
$^{9}$Department of Physics \& Astronomy, University of Southampton, Highfield, Southampton SO17 1BJ, UK\\
$^{10}$Cahill Center for Astronomy and Astrophysics, California Institute of Technology, Pasadena, CA 91125, USA\\
$^{11}$NASA Goddard Space Flight Center, Greenbelt, MD 20771, USA\\
$^{12}$NASA Postdoctoral Program Fellow\\
$^{13}$Department of Astronomy, Tsinghua University, Shuangqing Road 30, Beijing 100084 China\\
$^{14}$Tsinghua Center for Astrophysics, Tsinghua University, Shuangqing Road 30, Beijing 100084 China\\
$^{15}$Raman Research Institute, C. V. Raman Avenue, Sadashivanagar, Bangalore -- 560080, India\\
}

\date{Accepted XXX. Received YYY; in original form ZZZ}

\pubyear{2020}

\begin{document}
\label{firstpage}
\pagerange{\pageref{firstpage}--\pageref{lastpage}}
\maketitle

\begin{abstract}
The hard X-ray spectrum of magnetic cataclysmic variables can be modelled to provide a measurement of white dwarf mass.
This method is complementary to radial velocity measurements, which depend on the (typically rather uncertain) binary inclination.
%This method is independent of radial velocity measurements, which are often affected by uncertainties in binary inclination. 
Here we present results from a Legacy Survey of 19 magnetic cataclysmic variables with {\em NuSTAR}. We fit accretion column models to their 20--78 keV spectra and derive the white dwarf masses, finding a weighted average $\bar{M}_{\rm WD}=0.77\pm0.02$ \msun, with a standard deviation %about the weighted mean 
$\sigma=0.10$ \msun, when we include the masses derived from previous {\em NuSTAR} observations of seven additional magnetic cataclysmic variables. We find that the mass distribution of accreting magnetic white dwarfs is consistent with that of white dwarfs in non-magnetic cataclysmic variables. Both peak at a higher mass than the distributions of isolated white dwarfs and post-common-envelope binaries. We speculate as to why this might be the case, proposing that consequential angular momentum losses may play a role in accreting magnetic white dwarfs and/or that our knowledge of how the white dwarf mass changes over accretion--nova cycles may also be incomplete.

\end{abstract}


%%%%%%%%%%%%%%%%%%%%%%%%%%%%%%%%%%%%%%%%%%%%%%%%%%

%%%%%%%%%%%%%%%%% BODY OF PAPER %%%%%%%%%%%%%%%%%%

\section{Introduction}
\label{sec:intro}

Cataclysmic Variables (CVs) are binary systems in which a white dwarf (WD) accretes matter from a stellar companion via Roche lobe overflow \citep[for an in-depth review of CVs see][]{Warner-2003}. Magnetic CVs (mCVs) are a class of CVs in which the central WD has a strong magnetic field ($B\sim10^6-10^8$G), which disrupts the accretion disc and forces the accreted material to travel along the magnetic field lines on to the WD poles \citep[see reviews by][]{Cropper-1990,Patterson-1994}. Depending on the strength of the magnetic field, mCVs can be further divided into two subclasses. In Intermediate Polars (IPs), only the innermost regions of the disc are disrupted, typically leaving a residual outer disc which terminates at the magnetospheric radius (\rmag). It is at this point where matter begins to flow along the magnetic field lines in so-called `accretion curtains.' In polars, the magnetic field of the WD is strong enough such that matter flows along the field lines from the donor without forming a disc. 

Aside from the differing magnetic field strengths, polars and IPs can be observationally differentiated by the ratio of the WD spin period ($P_{\rm spin}$) to the binary orbital period ($P_{\rm orb}$). In polars, $P_{\rm spin}=P_{\rm orb}$, but in IPs $P_{\rm spin}<P_{\rm orb}$, usually $<<P_{\rm orb}$ \citep[see e.g.][]{Patterson-1994,Hellier-2014}. There are a small number\footnote{Five confirmed so far: BY\,Cam, V1500\,Cyg, CD\,Ind,  V1432\,Aql, and 1RXS J083842.1-282723 \citep{Halpern-2017,Rea-2017}.} of so-called `asynchronous' polars (APs) which exhibit the same properties as regular polars, but $P_{\rm spin}$ and $P_{\rm orb}$ differ by a factor of $\sim1$\%
\citep[see e.g.][]{Schmidt-1991,Littlefield-2015}.

Regardless of their subclass, mCVs are strong X-ray emitters \citep[see][for a review]{Mukai-2017}. Close to the surface of the WD, the in-falling material forms a standing shock, with typical temperatures of $kT\gtrsim10$ keV \citep{Aizu-1973}. As the gas in the post-shock region descends on to the surface of the WD and cools, it emits hard X-rays via optically thin thermal emission. It has been shown that the shock temperature is directly linked to the compactness of the WD \citep{Katz-1977,Rothschild-1981}. Thus, measuring the spectral turnover via hard X-ray spectroscopy of mCVs can be used to derive accurate WD masses. X-ray spectroscopy provides a method of measuring WD mass independent of radial velocity studies, which are often dominated by uncertainties in binary inclination \citep{Suleimanov-2005,Yuasa-2010,Suleimanov-2019}. An alternative X-ray spectroscopic method compares the fluxes from Fe K lines of different ionization states, to measure the post-shock temperature \citep{Fujimoto-1997,Ezuka-1999,Xu-2019}.

The derivation of WD mass is fundamental for quantative studies of individual objects. However, it is arguably even more important to know the WD mass distribution in order to understand the formation and evolution of CVs. The average mass of isolated WDs \citep[$\sim0.6M_{\odot}$;][]{Kepler-2016} is known to be lower than that of non-magnetic CVs \citep[$\sim0.8M_{\odot}$;][]{Zorotovic-2011}, though we note that for WDs within 100 pc, the mass distribution is apparently bimodal with peaks at 0.6 and 0.8 \msun\ \citep{Kilic-2018}. Furthermore, isolated {\em magnetic} WDs appear to be more massive on average \citep[$\bar{M}_{\rm WD}=0.784\pm0.047$ \msun;][]{Ferrario-2015} than their non-magnetized counterparts \citep[see also fig. 12 of][]{Ferrario-2020}.

In addition, comparing the WD mass distributions between non-magnetic and mCVs may be useful in testing theories of the origin of the magnetic field in WDs.
%Moreover, a comparison of WD mass distributions between non-magnetic and mCVs is of great interest in the context of the origin of the magnetic field in WDs. 
A leading scenario for the single magnetic WDs is that they are the results of mergers during the common envelope phase; mCVs are then understood to be the consequence of close interaction during the common envelope phase that end just short of merger (see e.g. \citealt{Ferrario-2015} but cf. \citealt{Belloni-2020}). Such a scenario could feasibly lead to a measurable difference between the mean masses of magnetic and non-magnetic WDs in CVs. Conversely, if magnetic and non-magnetic CVs share similar mass distributions, then one must question the evolution of WD mass once a binary becomes a CV, regardless of magnetic field strength. For example, the idea that CVs undergo a period of mass growth through accretion contradicts a number of existing theories of nova eruptions, which suggest that the amount of ejected mass is larger than the amount accreted \citep[e.g.][]{Prialnik-1995,Yaron-2005,Hillman-2020}.

Early (pre-2000) X-ray studies of WD masses in mCVs were limited to energies $\lesssim20$ keV that were probed by the X-ray observatories of the time \citep[e.g.][]{Cropper-1998,Cropper-1999}. The uncertainties were large, owing to the spectral cutoff (which is essential for mass determination) in mCVs usually occurring beyond 20 keV. However, with the inclusion of sensitive hard X-ray instruments on board satellites such as the {\em Rossi X-ray Timing Explorer} \citep[{\em RXTE};][]{Bradt-1993}, {\em Neil Gehrels Swift Observatory} \citep[{\em Swift};][]{Gehrels-2004}, {\em Suzaku} \citep{Mitsuda-2007} and the {\em International Gamma-Ray Astrophysics Laboratory} \citep[{\em INTEGRAL};][]{Winkler-2003}, mass measurements became more reliable and accurate. Studies with these instruments suggested that mCVs exhibit a similar mass distribution to that of their non-magnetic counterparts (see e.g. \citealt{Suleimanov-2005,Brunschweiger-2009,Yuasa-2010, Bernardini-2012}, as well as \citealt{deMartino-2020} and references therein for a review). Despite the improvement, these surveys still suffered from uncertain X-ray background \citep[which must be modelled rather than extracted for non-imaging instruments such as {\em Suzaku}'s Hard X-ray Detector;][]{Fukazawa-2009}.

% Early (pre-2000) X-ray studies of WD masses in mCVs were limited to energies $\lesssim20$ keV that were probed by the X-ray observatories of the time \citep[e.g.][]{Cropper-1998,Cropper-1999}. The uncertainties were large, owing to the spectral cutoff (which is essential for mass determination) in mCVs usually occurring beyond 20 keV. However, with the inclusion of sensitive hard X-ray instruments on board satellites such as the {\em Rossi X-ray Timing Explorer} \citep[{\em RXTE};][]{Bradt-1993}, {\em Neil Gehrels Swift Observatory} \citep[{\em Swift};][]{Gehrels-2004} and {\em Suzaku} \citep{Mitsuda-2007}, mass measurements became more reliable and accurate. Studies with these instruments suggested that mCVs exhibit a similar mass distribution to that of their non-magnetic counterparts \citep[see e.g.][]{Suleimanov-2005,Brunschweiger-2009,Yuasa-2010}. Despite the improvement, these surveys still suffered from uncertain X-ray background \citep[which must be modelled rather than extracted for non-imaging instruments such as {\em Suzaku}’s Hard X-ray Detector;][]{Fukazawa-2009}.

However, the emergence of the
{\em Nuclear Spectroscopic Telescope Array} \citep[\nus;][]{Harrison-2013} as the first telescope to be able to focus X-rays above 12 keV has brought about the ability to perform high angular-resolution imaging and spectroscopy in the hard X-ray regime. \nus\ is therefore the ideal instrument to perform a systematic survey of mCVs in order to efficiently measure the mass distribution of magnetic WDs. Mass measurements of IPs have been made with \nus\ previously \citep{Hailey-2016,Suleimanov-2016,Shaw-2018a,Suleimanov-2019}, but have  focused on a few sources at a time (for a total of 7 WD masses). In this work we present results from \nus\ observations of an additional 19 mCVs as part of the \nus\ Legacy Survey program.\footnote{\href{https://www.nustar.caltech.edu/page/legacy_surveys}{https://www.nustar.caltech.edu/page/legacy\_surveys}}

% The structure of this work is as follows:

\subsection{Modelling mCV masses}
\label{sec:models}

The standing shock in mCVs heats the infalling gas, which then cools via  optically thin thermal bremsstrahlung as it descends on to the surface of the WD. The hard X-ray continuum of mCVs, therefore, can be broadly modelled as a series of thermal bremsstrahlung components. However, the location of the shock, close to the WD surface, means that some X-ray emission will be directed towards the WD and reflected back towards the observer. Reflection modifies the underlying continuum with a Compton `hump' at $\sim10-30$ keV and neutral Fe-K emission at $6.4$ keV. When considering the whole \nus\ energy band, reflection has been found to be very important in modeling the X-ray spectrum of mCVs \citep{Mukai-2015}. Finally, the spectra of some mCVs may be affected by additional partial obscuration by the accretion curtains, even reaching the \nus\ band \citep[see e.g.][]{Done-1998,Cropper-1999,Shaw-2018a}.

For the mCVs discussed in this work, we use the IP mass model derived by \citet[][see these works for in-depth discussion of the model]{Suleimanov-2016,Suleimanov-2019}, which calculates a WD mass (\mwd) based on the temperature of the underlying continuum. We follow \citet{Suleimanov-2019} and refer to this as the `PSR' (post-shock region) model. \mwd\ is calculated assuming the \citet{Nauenberg-1972} mass-radius relation for cold WDs. The temperature of the shock, $kT_{\rm sh}$, depends on the velocity of infalling matter. Earlier models make the (often reasonable) assumption that the matter free-falls from infinity \citep[e.g.][]{Suleimanov-2005,Yuasa-2010}. In reality, and particularly in the case of IPs, free-fall begins at \rmag, where the accretion disc terminates. \rmag\ can %therefore 
be small enough in some cases that the accretion flow will %be accelerated to lower velocities, 
reach a velocity substantially smaller than the escape velocity,
leading to a lower value of $kT_{\rm sh}$. Assuming \rmag=$\infty$ in such cases would lead to an underestimation 
of the WD escape velocity, and thus
of \mwd. The PSR model therefore utilizes a two-parameter grid of hard X-ray spectra, with \mwd\ and \rmag\ (relative to the WD radius, $R_{\rm WD}$) as free parameters. \citet{Suleimanov-2019} introduced a slightly modified version of the model which also considers the height of the shock itself 
(as a shock height that is a significant fraction of the WD radius would also substantially reduce the escape velocity),
but this only becomes important for sources with a low local mass-accretion rate ($\lesssim1$ g s$^{-1}$ cm$^{-2}$). Given that the Legacy sample are all high luminosity \citep[$L\gtrsim10^{33}$ erg s$^{-1}$; see][see also Table \ref{tab:results}]{Suleimanov-2019}, we can assume that they have high local mass accretion rates too. However, we do discuss the effect that the shock height has on derived mass in Section \ref{subsec:shockheight}.

%No such sources are in our sample. {\color{red}{Double check this!  --and explain why we're sure...}}
%\blue{SV: We cannot estimate local mass accretion rates directly. We can only suggest that the high luminous IPs have the high local mass accretion rates as well. Therefore, I suggest to add information about the observed fluxes and the IPs luminosities in Table 2. All the investigated here IPs have the high enough luminosities in according to BAT data.}

When applied to the hard X-ray spectrum of an mCV, the PSR model defines a curve in the \mwd---\rmag\ plane. One can then derive another curve in the same plane by measuring a break in the power spectrum of the source light curve. The concept here is that the accretion disk generates noise at frequencies related to the orbital frequency, so the power spectrum cuts off at high frequency, where the magnetosphere truncates the inner accretion disk.  \citet{Revnivtsev-2009,Revnivtsev-2011} showed that the break frequency $\nu_{\rm b}$ corresponds to the Keplerian frequency at \rmag\ according to

\begin{equation}
    \nu_{\rm b} = \sqrt{\frac{G M_{\rm WD}}{2\pi R^{3}_{\rm m}}}.
    \label{eq:break}
\end{equation}

\noindent The intersection of the two curves then allows us to derive \mwd\ and \rmag\ (see Fig. \ref{fig:FO_Aqr} for an example for a Legacy survey target). \citet{Suleimanov-2019} applied this technique to 5 IPs that exhibited power spectrum breaks, finding that \mwd\ only changes significantly if $R_{\rm m}\lesssim4$ \rwd, as in the cases of GK\,Per (in outburst) and EX\,Hya.

\section{Observations and Analysis}
\label{sec:obs}

%add MJD /UTC to table?
\setlength{\tabcolsep}{5pt}
\begin{table*}
    \centering
    \caption{Summary of \nus\ observations}
    \begin{tabular}{l|c|c|c|c}
    \hline\hline
         Source & ObsID & Start Date/Time  & MJD$^{a}$ & Exposure time \\
         & & (UTC) & &(ks) \\
         \hline
         \vspace{2pt}1RXS\,J052523.2$+$241331 & 30460020002 &  2019 Mar 21 13:31:09 & 58563.56 & 58.8 \\
         \vspace{2pt}V515\,And & 30460019002 & 2019 Mar 09 20:51:09 & 58551.87 & 62.2\\         
         \vspace{2pt}V1432\,Aql$^*$ & 30460004002 & 2018 Apr 05 04:51:09 & 58213.20 & 27.2\\
         \vspace{2pt}FO\,Aqr & 30460002002 & 2018 Apr 16 02:01:09 & 58224.08 & 25.6\\
         \vspace{2pt}V405\,Aur & 30460007002 & 2017 Nov 08 20:01:09 & 58065.83 & 38.3\\
         \vspace{2pt}BY\,Cam$^*$ & 30460010002 & 2018 Nov 12 15:21:09 & 58434.64 & 33.0\\
         \vspace{2pt}BG\,CMi & 30460018002 & 2018 Oct 09 19:46:09 & 58400.82 & 40.4 \\
         \vspace{2pt}V2069\,Cyg & 30460023002 & 2019 Jun 27 23:46:09 & 58661.99 & 67.4 \\
         \vspace{2pt}PQ\,Gem & 30460009002 & 2019 Mar 27 14:36:09 & 58569.61 & 42.0\\
         \vspace{2pt}V2400\,Oph & 30460003002 & 2019 Mar 07 18:31:09 & 58549.77 & 27.0\\
         \vspace{2pt}AO\,Psc & 30460008002 & 2018 Jun 29 04:21:09 & 58298.18  & 37.4\\
         \vspace{2pt}V667\,Pup & 30460012002 & 2019 May 20 21:36:09 & 58623.90 & 39.4\\
         \vspace{2pt}V1062\,Tau & 30460015002 & 2020 Mar 17 01:31:09 & 58925.06 & 31.5\\
         \vspace{2pt} ...  & 30460015004 & 2020 Mar 17 18:31:49 &58925.77  & 30.4\\
         \vspace{2pt}EI\,UMa & 30460011002 & 2019 Mar 20 18:21:09 & 58562.76 & 35.0 \\
         \vspace{2pt}IGR\,J08390$-$4833 & 30460025002 & 2018 Feb 09 08:01:09 & 58158.33 & 55.2\\
         \vspace{2pt}IGR\,J15094$-$6649 & 30460013002 & 2018 Jul 19 23:01:09 & 58318.96 & 41.3\\
         \vspace{2pt}IGR\,J16547$-$1916 & 30460016002 & 2019 Mar 16 05:16:09 & 58558.22 & 44.6 \\
         \vspace{2pt}IGR\,J17195$-$4100 & 30460005002 & 2018 Oct 25 22:56:09 & 58416.96 & 29.5 \\
         \vspace{2pt}RX\,J2133.7$+$5107 & 30460001002 & 2018 Feb 23 12:51:09 & 58172.54 & 26.2\\
         \hline
    \end{tabular}
    \label{tab:observations}\\
    \small\raggedright$^*$Asynchronous Polar\\
    \small\raggedright$^{a}$ MJD at the start of the observation
\end{table*}

In this work we utilize observations of 17 IPs and two APs from the {\em NuSTAR} Legacy survey of mCVs, spanning a period of $\sim2.5$yr. Individual targets, observation dates and their exposure times are detailed in Table \ref{tab:observations}. There were four additional targets, observed as part of the Legacy programme, that we do not include in the analysis for various reasons. IGR\,J14536$-$5522 is a polar that was not detected in a 46.0 ks \nus\ observation, and was likely in a low state due to a reduced mass-accretion rate \citep[commonly seen in polars;][]{Ramsay-2004}. YY/DO\,Dra is an IP that was not detected in a 55.4 ks \nus\ observation. Low states are much rarer in IPs than in polars \citep[see e.g.][]{Kennedy-2017}, but a {\em Swift} observation, quasi-simultaneous with \nus, confirms the low state of YY/DO\,Dra. XY\,Ari is an IP that was observed by \nus, but the observation was interrupted by a high priority target of opportunity observation and was never completed. The 9.3 ks of data that do exist are not enough to constrain a mass using the methodology we describe below. RX\,J2015.6$+$3711 is a CV of uncertain classification, but has been suggested to be an IP \citep[][]{CotiZelati-2016}. A 59.6 ks \nus\ observation does not allow us to constrain a mass using the methodology we describe below as the source is a factor $\sim6$ fainter than the reported {\em Swift}/Burst Alert Telescope \citep[BAT;][]{Barthelmy-2005} 70 month catalogue flux \citep{Mukai-2017}, raising the possibility that, like DO\,Dra, RX\,J2015.6$+$3711 is also in a low state.

Three Legacy targets (FO\,Aqr, V405 Aur and RX\,J2133.7$+$5107) have been discussed by \citet{Suleimanov-2019} alongside the 7 IPs previously observed with {\em NuSTAR}, but we re-reduce and analyse the data for those three in this study. We do not re-reduce the observations of the 7 previously observed IPs, instead choosing to combine our Legacy results with the published results from those sources \citep[see][and references therein for detailed analyses of the non-Legacy data]{Suleimanov-2019}.

We reduced the data using the \nus\ data analysis software (NuSTARDAS) v1.8.0 packaged with {\sc heasoft} v6.26.1. The exception to this is the observation of V1062\,Tau, which took place on 2020 March 17, and thus required NuSTARDAS v1.9.2 (packaged with {\sc heasoft} v6.27.2) in order to account for the adjustment of \nus's onboard laser metrology system\footnote{\href{https://heasarc.gsfc.nasa.gov/docs/nustar/analysis/}{https://heasarc.gsfc.nasa.gov/docs/nustar/analysis/}}. Data taken prior to 2020 March 17 are unaffected by this change so reprocessing was not necessary.

We used the {\tt nupipeline} task to perform standard data processing, including filtering for high levels of background during the telescope's passage through the South Atlantic Anomaly and generation of exposure maps. We extracted spectra and (10s binned) light curves from the resultant cleaned event files (from both \nus\ focal plane modules; FPMA and FPMB) using the {\tt nuproducts} task. Source spectra and light curves were extracted from a circular region of radius ranging from $30-70$\arcsec. %, depending on source brightness.
The background was typically extracted from a 70\arcsec circular region in the opposite corner of the same chip that the source lay on. However, the observation of the IP 1RXS\,J052523.2$+$241331 was badly affected by photons from a nearby source that bypassed the telescope optics \citep["stray light",][]{Madsen-2017}. In both FPMs, the source fell on the region of the detector containing the stray light, and we extracted the background spectrum from a 50\arcsec region that included the stray light photons. %We also took care to avoid areas of stray light on the chip when selecting background regions.

We grouped the spectra such that each spectral bin had a signal-to-noise ratio S/N=3 using the {\sc heasoft} task {\tt grppha}. Light curves from FPMA and FPMB were co-added with the {\sc heasoft} tool {\tt lcmath} and then corrected to the solar system barycentre with {\tt barycorr}.

\begin{figure}
    \centering
    \begin{subfigure}[b]{0.45\textwidth}
        \includegraphics[width=\textwidth]{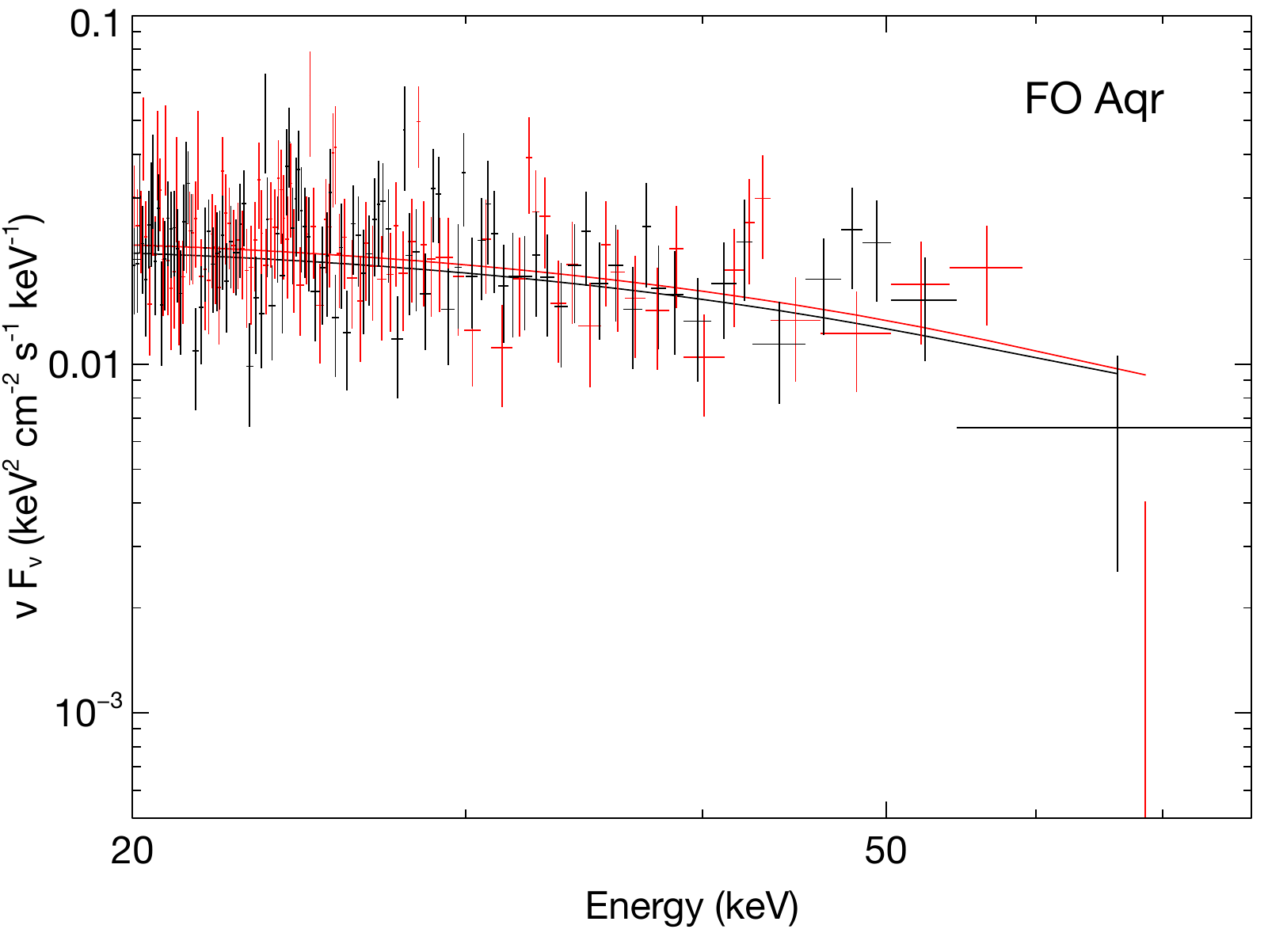}
    % \vspace{-5mm}
    \end{subfigure}
    % \vspace{-3mm}
    \begin{subfigure}[b]{0.45\textwidth}
        \includegraphics[width=\textwidth]{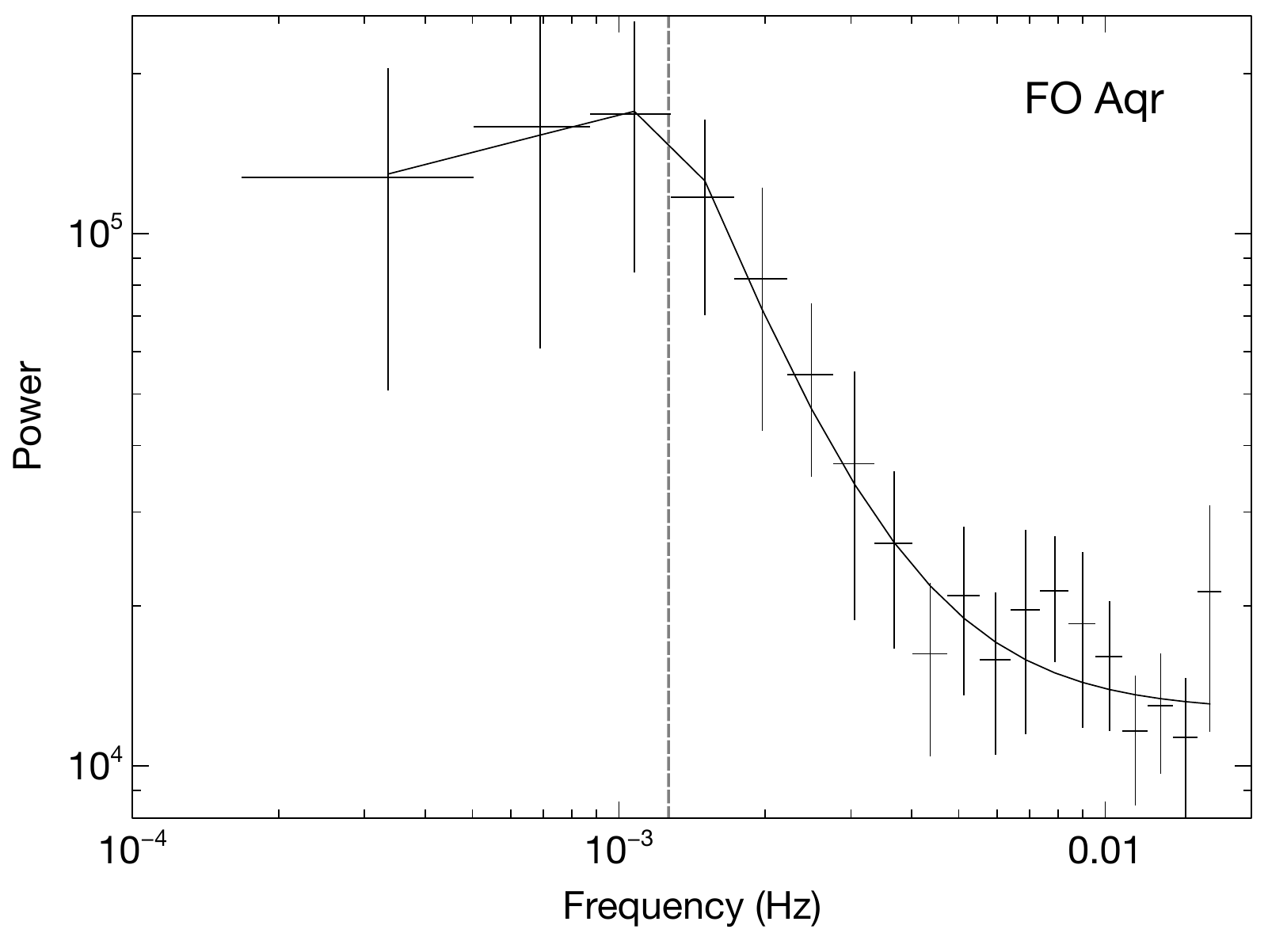}
    \end{subfigure}
    % \vspace{-3mm}
    \begin{subfigure}[b]{0.5\textwidth}
        \includegraphics[width=\textwidth]{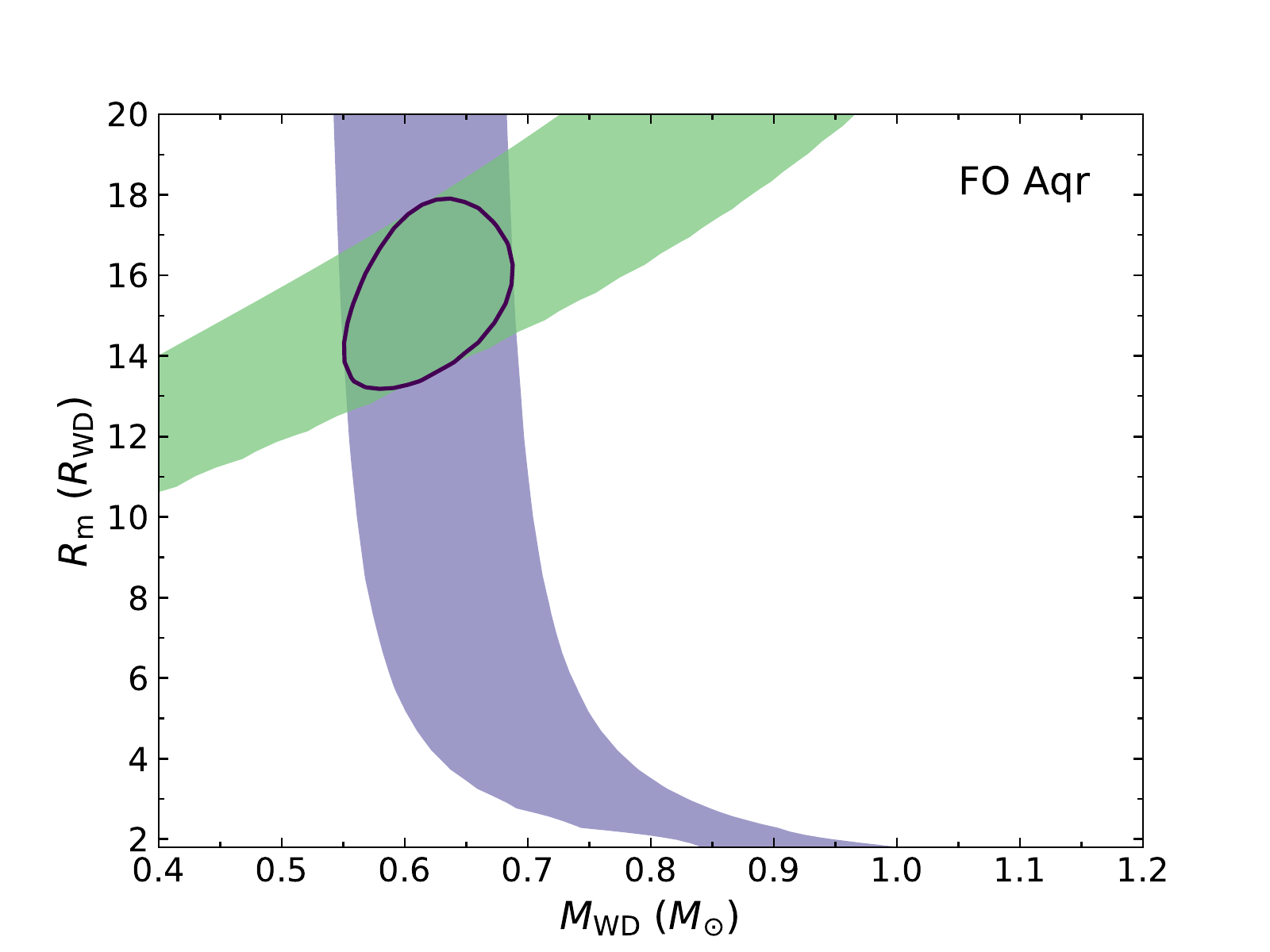}
    \end{subfigure}
    \caption{{\em Upper Panel}: \nus\ FPMA (black) and FPMB (red) spectra of FO\,Aqr, fit with the \citet{Suleimanov-2019} PSR model and plotted unfolded in $\nu$F$_{\nu}$ space. {\em Middle Panel}: Aperiodic power spectrum of the \nus\ light curve of FO\,Aqr, fit with a broken power law. The power spectrum shows a break at $\nu_{\rm b}=1.3\times10^{-3}$ Hz (indicated by the vertical dashed line), which can be linked to the \rmag\ and \mwd\ parameters in the (energy) spectral model in {\sc xspec}. {\em Lower Panel}: 90\% confidence contours in the \mwd---\rmag\ plane calculated by fitting the power spectrum (green) and the energy spectrum (purple) separately. The solid closed curve represents the 90\% confidence region obtained by fitting the power and energy spectra together. The best-fit parameters for FO\,Aqr are presented in Table \ref{tab:results}.}
    \label{fig:FO_Aqr}
\end{figure}

The break in IP power spectra discussed in Section \ref{sec:models} is easiest to detect if the periodic variability from WD spin is removed from the light curve first. To do this we followed a similar method as  \citet{Suleimanov-2019}. We first split the \nus\ light curve into segments, the number of which was dependent on the length of the light curve, the source count rate and the WD \pspin. We folded each segment on the known \pspin\ from the  literature. Using the same folding parameters, we calculated the pulse phase for each light curve segment time stamp and subtracted the expected rate from the observed rate. The power spectra of the resultant aperiodic light curves were then calculated using the {\tt stingray} {\sc python} library, a suite of tools dedicated to time series analysis  \citep{Huppenkothen-2016,Huppenkothen-2019}. We note that the above analysis only applies to the 17 IPs in our sample, as the APs do not have a disc. We detect a break in only one of our sample, FO\, Aqr at $\nu_{\rm b}=1.3\times10^{-3}$ Hz, confirming the findings of \citet{Suleimanov-2019}.\footnote{The frequency range in which we searched for breaks in the power spectra was dependent on the length of the light curve segments and the WD \pspin\ but typically ranged from $\sim3\times10^{-4}$ -- $0.02$ Hz.} For this source we used the tool {\sc flx2xsp} to convert the aperiodic power spectrum into a format readable by the X-ray spectral fitting package {\sc xspec} \citep{Arnaud-1996}. It is important to note that the non-detection of a break in the power spectra of our target IPs does not have strong implications for the derived \mwd. The bottom panel of Fig. \ref{fig:FO_Aqr} shows that \mwd\ is insensitive to changes in \rmag\ unless it is very close ($\lesssim4$\rwd) to the WD, in which case \mwd\ would increase. We might expect this in sources with a higher than typical mass accretion rate due to e.g. a dwarf nova outburst, where the magnetosphere is compressed \citep[see e.g.][and their discussion of GK Per]{Suleimanov-2019}, but not for the majority of IPs.

To derive \mwd\ for each source, we fit its X-ray spectrum with the \citet{Suleimanov-2019} PSR model. As discussed in Section \ref{sec:models}, the X-ray spectrum of mCVs is often complicated by the presence of a combination of partial covering and reflection components. When considering the whole NuSTAR band (3--78 keV), the derivation of \mwd\ then relies heavily on the choice of reflection and absorption models to account for these effects, and no robust description of both such components exists in the context of mCVs \citep[see discussions by][]{Suleimanov-2016,Suleimanov-2019}. Furthermore, in the PSR model, \mwd\ is mostly driven by the turnover of the spectrum at high energies. We therefore choose to restrict our fitted energy range to 20--78 keV in order to minimize the contributions from reflection and partial covering effects to the overall spectrum. This is not an unusual approach \citep[see discussions by e.g.][]{Suleimanov-2016,Hailey-2016,Suleimanov-2019}, and it allows us to derive \mwd\ without having to consider the multitude of complex effects that dominate the X-ray spectrum below 20 keV. There is one exception: for 1RXS\,J052523.2$+$241331, we instead restricted the fit to 20--50 keV, to minimize the contribution by stray light photons, which dominate the spectrum beyond $>50$ keV.

%To minimize the contributions from reflection and partial covering effects, we restrict the fitted energy range to 20--78 keV \citep[see discussions by][]{Suleimanov-2016,Hailey-2016,Suleimanov-2019} for most of the sources. However, for 1RXS\,J052523.2$+$241331, we restricted the fit to 20--50 keV, to minimize the contribution by stray light photons, which dominate the spectrum beyond $>50$ keV. For the unique case of FO\,Aqr, where we detect a break in the power spectrum, we are able to link the break frequency to the \rmag\ parameter in the model using Equation \ref{eq:break} and fit the energy and power spectra simultaneously (see Fig. \ref{fig:FO_Aqr}). 

For the unique case of FO\,Aqr, where we detect a break in the power spectrum, we are able to link the break frequency to the \rmag\ parameter in the model using Equation \ref{eq:break} and fit the energy and power spectra simultaneously (see Fig. \ref{fig:FO_Aqr}). For the other IPs, we assume that the WD is close to %corotation 
spin equilibrium; that is, that accreting matter has the same angular velocity as the WD. If this is not the case, there will be a torque trying to slow or speed the WD's rotation. For IPs that have accreted persistently for a long time, we may assume that the WD has come into spin equilibrium \citep[see e.g.][]{Patterson-2020}. 
%such that
In this case, the Keplerian velocity of accreting material at \rmag\ will match the WD's spin; 

\begin{equation}
    R_{\rm m} \simeq R_{\rm co} = \left(\frac{G M_{\rm WD} P^2_{\rm spin}}{4 \pi^2}\right)
    \label{eq:Rco}
\end{equation}

\noindent where \rco\ is the corotation radius. This is a reasonable assumption for all the IPs in our sample, which are persistent sources that show no signs of transient outbursts in public all-sky monitoring data from the {\em Swift}/BAT\footnote{\href{https://swift.gsfc.nasa.gov/results/transients/}{https://swift.gsfc.nasa.gov/results/transients/}} and {\em Monitor of All-Sky X-ray Image} \citep[{\em MAXI};][]{Matsuoka-2009}\footnote{\href{http://maxi.riken.jp/top/slist.html}{http://maxi.riken.jp/top/slist.html}}.
% We show in Fig. \ref{fig:kT_vs_mass} that the systematic uncertainties that would arise as a result of a poor understanding of \rmag are much smaller than the statistical uncertainties.

% \begin{figure}
%     \centering
%     \includegraphics[width=0.45\textwidth]{Figures/kT_vs_mass_SN.pdf}
%     \caption{Comparison of the derived WD masses with the temperatures derived from a best-fit bremsstrahlung model. The solid red line represents relation between $kT$ and \mwd\ as derived by \citet{Suleimanov-2016}, assuming a magnetospheric radius equal to the average derived for our sample $\bar R_{\rm m}=17.0$. The upper dashed red line is the same relation but for $R_{\rm m}=7.4$ (the smallest \rmag\ in our IP sample) and the lower dashed red line is the same relation but for \rmag=$\infty$.}
%     \label{fig:kT_vs_mass}
% \end{figure}

To include the spin equilibrium assumption in our {\sc xspec} spectral fits, we set the \rmag\ parameter to be a function of the \mwd\ parameter using Equation \ref{eq:Rco}. For the APs, we assume %\rmag$=\infty$ (which is entered into {\sc xspec} as
$R_{\rm m}=1000 R_{\rm WD}$, which is equivalent to $R_{\rm m}=\infty$ in the {\sc xspec} model. For all observations, we fit the FPMA \& FBMB spectra simultaneously, with the cross-normalisation between the two instruments accounted for by a constant. All {\sc xspec} fits utilised $\chi^2$ as the fitting statistic, and all uncertainties presented in this work are given at $90\%$ confidence unless otherwise stated.

\section{Results and Discussion}
\label{sec:discussion}

\setlength{\tabcolsep}{5pt}
\begin{table*}
    \centering
    \caption{Results of the \nus\ Legacy Survey. The values of \rmag\ and \mwd\ are extracted from the best-fit PSR model and $kT_{\rm bremss}$ is extracted from a single temperature bremmstrahlung fit to the spectra ({\tt bremss} in {\sc xspec}) and is not intended to be a full treatment of the shock temperature \citep[see e.g.][]{Mukai-2015}.}
    \begin{tabular}{l|c|c|c|c|c|c}
    \hline\hline
         Source & \rmag\  & \mwd\ & $kT_{\rm bremss}$  & $F^a$  & $d^b$ & $L^c$  \\
          & (\rwd) & (\msun) & (keV) & ($10^{-10}$ erg cm$^{-2}$ s$^{-1}$) & (pc) & ($10^{33}$ erg s$^{-1}$) \\
         \hline
         \vspace{2pt}1RXS\,J052523.2+241331$^{d}$ & $7.4^*$& $0.81^{+0.13}_{-0.10}$ & $19.6^{+5.2}_{-3.6}$ & $0.55^{+0.09}_{-0.07}$ & $1888^{+363}_{-266}$ & $23.5^{+9.9}_{-7.2}$\\ 
         \vspace{2pt}V515\,And & $10.7^*$ & $0.73\pm0.06$ & $18.2^{+2.3}_{-2.0}$ & $0.70^{+0.08}_{-0.06}$ &$978^{+46}_{-42}$ & $8.0^{+0.12}_{-0.10}$\\
         \vspace{2pt}V1432\,Aql & $\infty^\dagger$ & $0.76^{+0.09}_{-0.08}$ & $20.8^{+3.9}_{-3.1}$ & $0.85^{+0.12}_{-0.09}$ & $456^{+10}_{-9}$ & $2.1^{+0.3}_{-0.2}$ \\
         \vspace{2pt}FO\, Aqr & $15.0^{+2.1}_{-1.5}$ & $0.61^{+0.06}_{-0.05}$ & $14.8^{+1.8}_{-1.7}$ & $1.41^{+0.20}_{-0.17}$ & $518^{+14}_{-13}$ & $4.5^{+0.7}_{-0.6}$\\
         \vspace{2pt}V405\,Aur & $12.3^*$ & $0.75^{+0.07}_{-0.06}$ & $19.1^{2.6}_{-2.2}$ & $0.95^{+0.11}_{-0.09}$ & $662^{+14}_{-13}$ & $5.0^{+0.6}_{-0.5}$ \\
         \vspace{2pt}BY\,Cam & $\infty^\dagger$ & $0.76\pm0.06$ & $21.0^{+2.4}_{-2.0}$ & $1.49^{+0.13}_{-0.11}$ & $270\pm2$ & $1.3\pm0.1$\\
         \vspace{2pt}BG\,CMi & $18.2^*$& $0.78^{+0.08}_{-0.07}$ & $20.7^{+3.1}_{-2.6}$ & $0.74^{+0.09}_{-0.07}$ & $966^{+55}_{-50}$ & $8.3^{+1.3}_{-1.2}$\\
         \vspace{2pt}V2069\,Cyg & $14.7^*$ & $0.73^{+0.09}_{-0.08}$ & $18.5^{+2.8}_{-3.6}$ & $0.46^{+0.07}_{-0.06}$ & $1140^{+43}_{-40}$ & $7.2^{+1.3}_{-1.0}$\\
         \vspace{2pt}PQ\,Gem & $15.3^*$ & $0.71^{+0.06}_{-0.05}$ & $18.0^{+2.3}_{-1.9}$ & $0.94^{+0.11}_{-0.09}$ & $750^{+21}_{-20}$ & $6.3^{+0.8}_{-0.7}$\\
         \vspace{2pt}V2400\,Oph & $15.5^*$ & $0.67^{+0.06}_{-0.05}$ & $16.6^{+2.0}_{-1.7}$ & $1.59^{+0.19}_{-0.16}$ & $701^{+17}_{-16}$ & $9.3^{+1.2}_{-1.0}$ \\
         \vspace{2pt}AO\,Psc & $11.6^*$ & $0.55\pm0.05$ & $12.4^{+1.7}_{-1.4}$ & $1.48^{+0.28}_{-0.22}$ & $488^{+11}_{-10}$ & $4.2^{+0.8}_{-0.6}$\\
         \vspace{2pt}V667\,Pup & $13.4^*$ & $0.83^{+0.11}_{-0.08}$ & $22.8^{+4.3}_{-3.4}$ & $0.63^{+0.07}_{-0.06}$ & $798^{+51}_{-46}$ & $5.1^{+0.9}_{-0.8}$\\
         \vspace{2pt}V1062\,Tau & $43.0^*$ & $0.72^{+0.07}_{-0.06}$ & $18.7^{+2.4}_{-2.0}$ & $0.71^{+0.08}_{-0.07}$  & $1512^{+209}_{-165}$ & $19.5^{+5.8}_{-4.7}$\\
         \vspace{2pt}EI\,UMa & $19.4^*$& $0.91^{+0.15}_{-0.13}$ & $26.0^{+8.1}_{-5.6}$ & $0.39^{+0.06}_{-0.05}$ & $1095^{+47}_{-43}$ & $5.6^{+1.0}_{-0.8}$\\
         \vspace{2pt}IGR\,J08390$-$4833 & $26.4^*$& $0.81^{+0.13}_{-0.11}$ & $12.1^{+5.8}_{4.2}$ & $0.30^{+0.05}_{-0.04}$ & $2064^{+311}_{-240}$ & $15.4^{+5.4}_{-4.1}$\\
         \vspace{2pt}IGR\,J15094$-$6649 & $15.4^*$ & $0.73\pm0.06$ & $18.5^{+2.4}_{-2.0}$ & $0.97^{+0.11}_{-0.09}$ & $1127^{+37}_{-35}$ & $14.8^{+1.9}_{-1.7}$\\
         \vspace{2pt}IGR\,J16547$-$1916 & $12.2^*$& $0.74^{+0.09}_{-0.08}$ & $18.9^{+3.4}_{-2.7}$ & $0.58^{+0.09}_{-0.07}$ & $1066^{+61}_{-54}$ & $7.8^{+1.5}_{-1.2}$\\
         \vspace{2pt}IGR\,J17195$-$4100 & $21.8^*$& $0.84^{+0.08}_{-0.07}$ & $22.9^{+3.6}_{-2.9}$ & $0.97^{+0.10}_{-0.08}$ & $643^{+17}_{-16}$ & $4.8^{+0.6}_{-0.5}$\\
         \vspace{2pt}RX\,J2133.7+5107 & $17.5^*$& $0.96^{+0.08}_{-0.07}$ & $28.2^{+4.6}_{-3.7}$ & $1.06^{+0.08}_{-0.07}$ & $1325^{+48}_{-45}$ & $22.2^{+2.4}_{-2.2}$\\
         \hline
    \end{tabular}
    \label{tab:results}\\
    \small\raggedright$^a$ Best-fit model flux (unabsorbed) from the PSR model in the 0.1--100 keV range, calculated using the {\tt cflux} model in {\sc xspec} (we choose this range for easy comparisons with \citet{Suleimanov-2019})\\
    \small\raggedright$^b$ Distance from {\em Gaia} DR2 \citep{Bailer-Jones-2018}\\
    \small\raggedright$^c$ Luminosity calculated from best-fit model flux of the PSR model in 0.1--100 keV range \\
    \small\raggedright$^{d}$Spectrum was fit in the 20--50 keV range due to stray light\\
    \small\raggedright$^*$ $R_{\rm m}=R_{\rm co}$ assumed\\ 
    \small\raggedright$^\dagger$ $R_{\rm m}=1000 R_{\rm WD}$ assumed in {\sc xspec}\\
\end{table*}

\subsection{Comparisons with other WD distributions}
\label{subsec:comparisons}

The measured masses from our 19 Legacy targets are listed in Table \ref{tab:results} and the spectra are plotted for reference in Fig. \ref{fig:spec_figs}. For the Legacy sample, we calculate a weighted average $\bar{M}_{\rm WD}=0.72$, which increases to $\bar{M}_{\rm WD}=0.77$ if we include the 7 IPs previously observed with \nus\ \citep{Suleimanov-2019}. We refer to the Legacy+\citet{Suleimanov-2019} sample as the "full \nus\ sample." To estimate uncertainties on $\bar{M}_{\rm WD}$, we first calculate the standard deviation of the sample around the weighted average and find $\sigma=0.09$ \msun\ ($\sigma=0.10$ \msun) for the Legacy (full \nus) samples. In both cases, this value is larger than what is implied by the numerical error in the weighted mean (0.01 and 0.006 \msun, respectively) after correcting for the population size. This can occur if the errors of individual data points are underestimated. To quantitatively address this, we calculated 10000 weighted averages from a randomly selected sample of the Legacy only, or Legacy+\citet{Suleimanov-2019} mCVs (bootstrap-with-replacement) and measured the 68\% confidence interval. We find this to be $0.02$ \msun\ for both the Legacy  and full \nus\ samples. The correction of this for population size agrees with the calculated values of $\sigma$ above. Since the mean of the 10000 bootstrapped weighted averages agrees with the weighted average of the full sample, we find no evidence of bias in our masses. We thus quote weighted averages as follows: $\bar{M}_{\rm WD}=0.72\pm0.02$ \msun\ for the Legacy only sample, and $\bar{M}_{\rm WD}=0.77\pm0.02$ \msun\ for the full \nus\ sample.

The average mass for the full \nus\ mCV sample is consistent with
%close to
that of IPs obtained with non-imaging telescopes \citep[$0.88\pm0.25$ and $0.86\pm0.07$ \msun;][respectively]{Yuasa-2010,Bernardini-2012}, and with a combination of non-imaging and imaging telescopes \citep[$0.84\pm0.17$ \msun;][]{deMartino-2020}, though slightly lower. We note here that the results of \citet{Bernardini-2012} may be biased towards higher masses as the majority of their sample consists of sources that were discovered by {\em INTEGRAL}, a hard X-ray telescope. \citet{Yuasa-2010} also note that their sample of 16 of the brightest IPs may be biased towards higher masses. We discuss potential selection biases of our sample in Section \ref{subsec:sel_effects}.

%The average mass for the full \nus\ mCV sample is close to that of non-magnetic CVs \citep[$0.83\pm0.23$ \msun;][]{Zorotovic-2011}, as well as that of IPs obtained with non-imaging telescopes \citep[$0.88\pm0.25$, $0.86\pm0.07$ and $0.79\pm0.16$ \msun;][respectively]{Yuasa-2010,Bernardini-2012, Suleimanov-2019}. We note here that the results of \citet{Bernardini-2012} are likely biased towards higher masses as the majority of their sample consists of sources discovered by {\em INTEGRAL}, a hard X-ray telescope. \citet{Yuasa-2010} also note that their sample of 16, flux-selected IPs may be biased towards higher masses.

\begin{figure}
    \centering
    \includegraphics[width=0.5\textwidth]{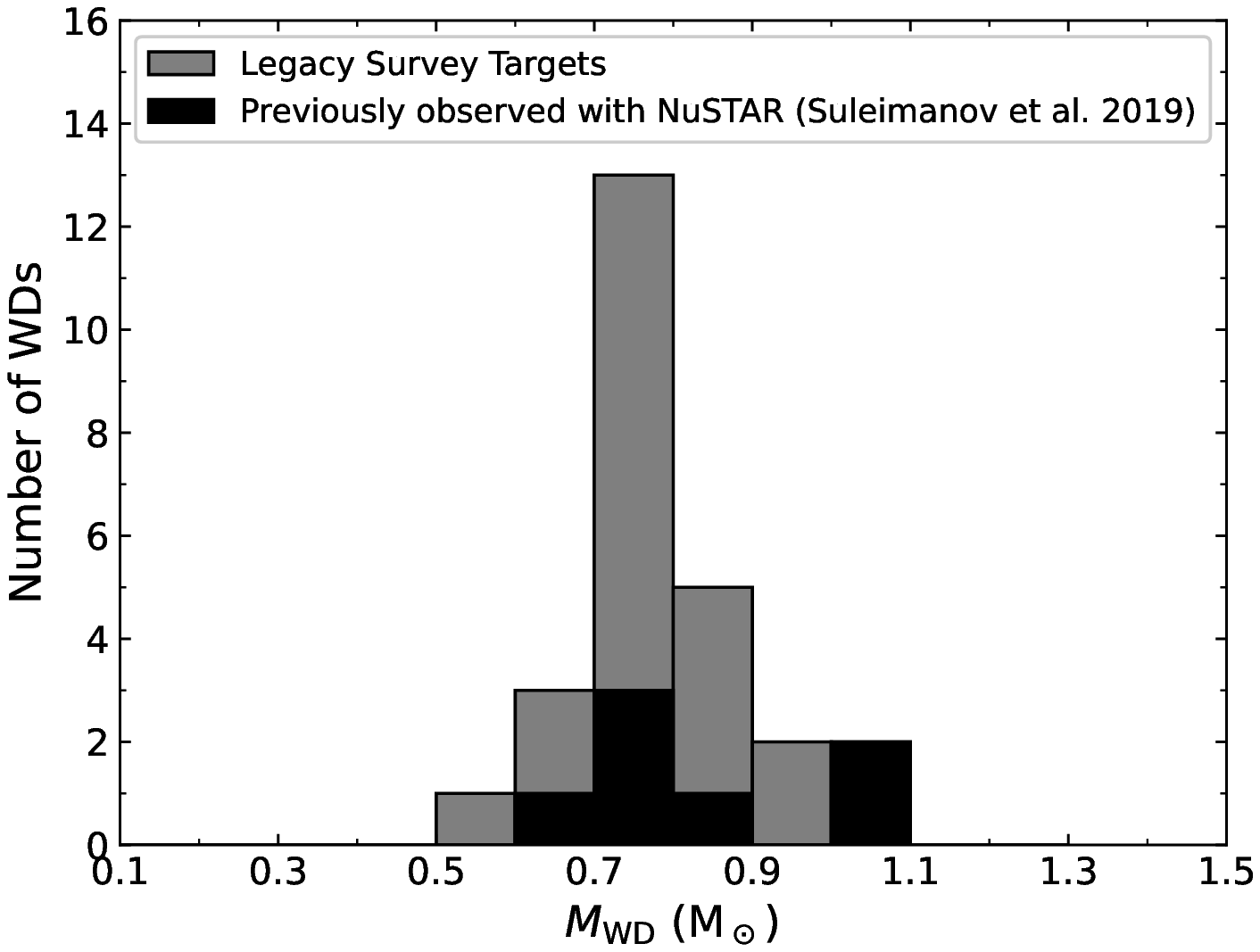}
    \caption{Stacked histogram of the mass distribution of WDs in mCVs. %Masses are weighted by the $1\sigma$ uncertainties on \mwd\ (see Section \ref{subsec:comparisons}).
    Legacy targets from this work are in grey, whilst masses derived from previous \nus\ observations \citep{Suleimanov-2019} are in black.}
    \label{fig:mass_hist}
\end{figure}

\begin{figure}
    \centering
    \begin{subfigure}[b]{0.5\textwidth}
        \includegraphics[width=\textwidth]{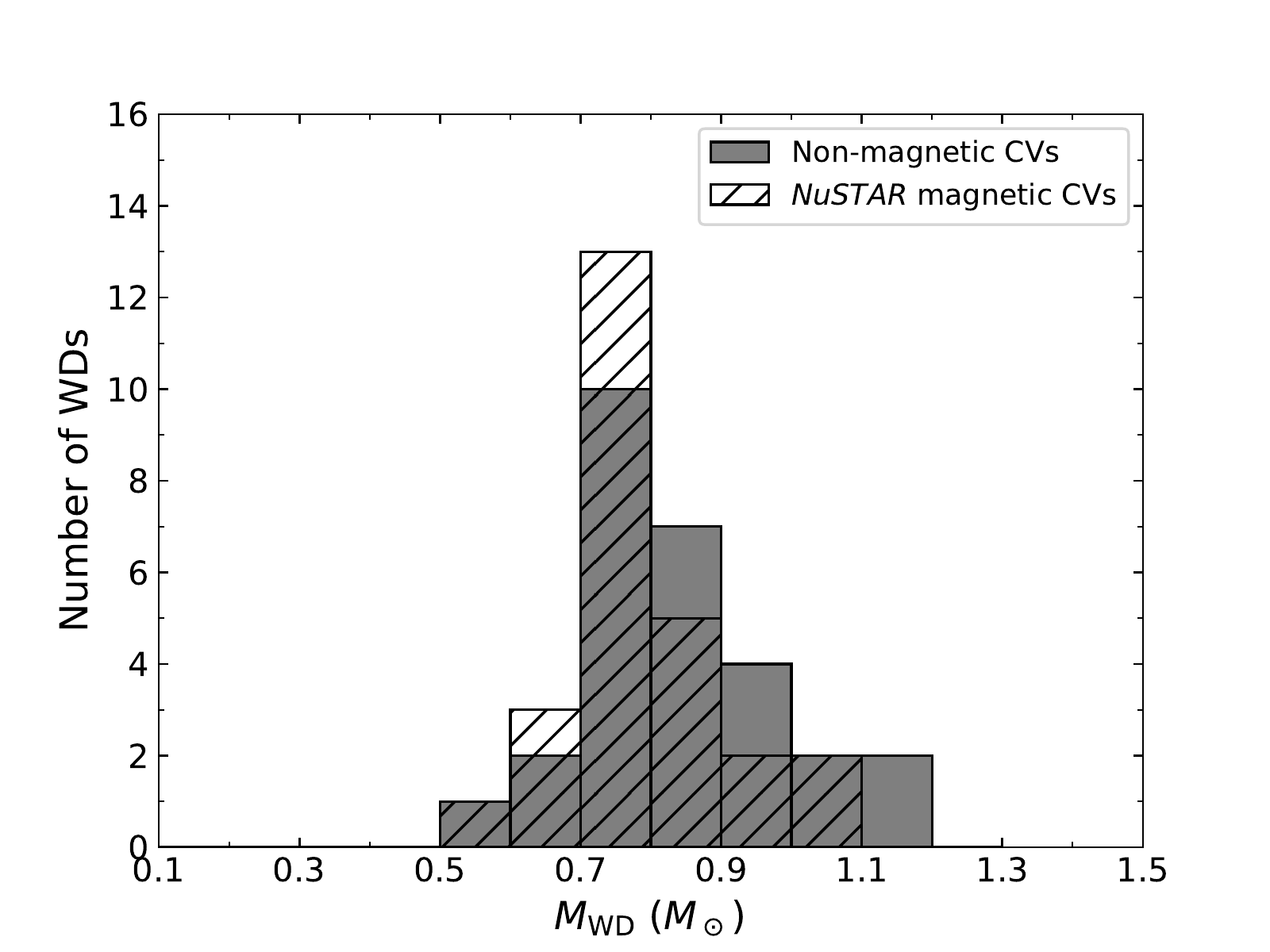}
    % \vspace{-5mm}
    \end{subfigure}
    % \vspace{-3mm}
    \begin{subfigure}[b]{0.5\textwidth}
        \includegraphics[width=\textwidth]{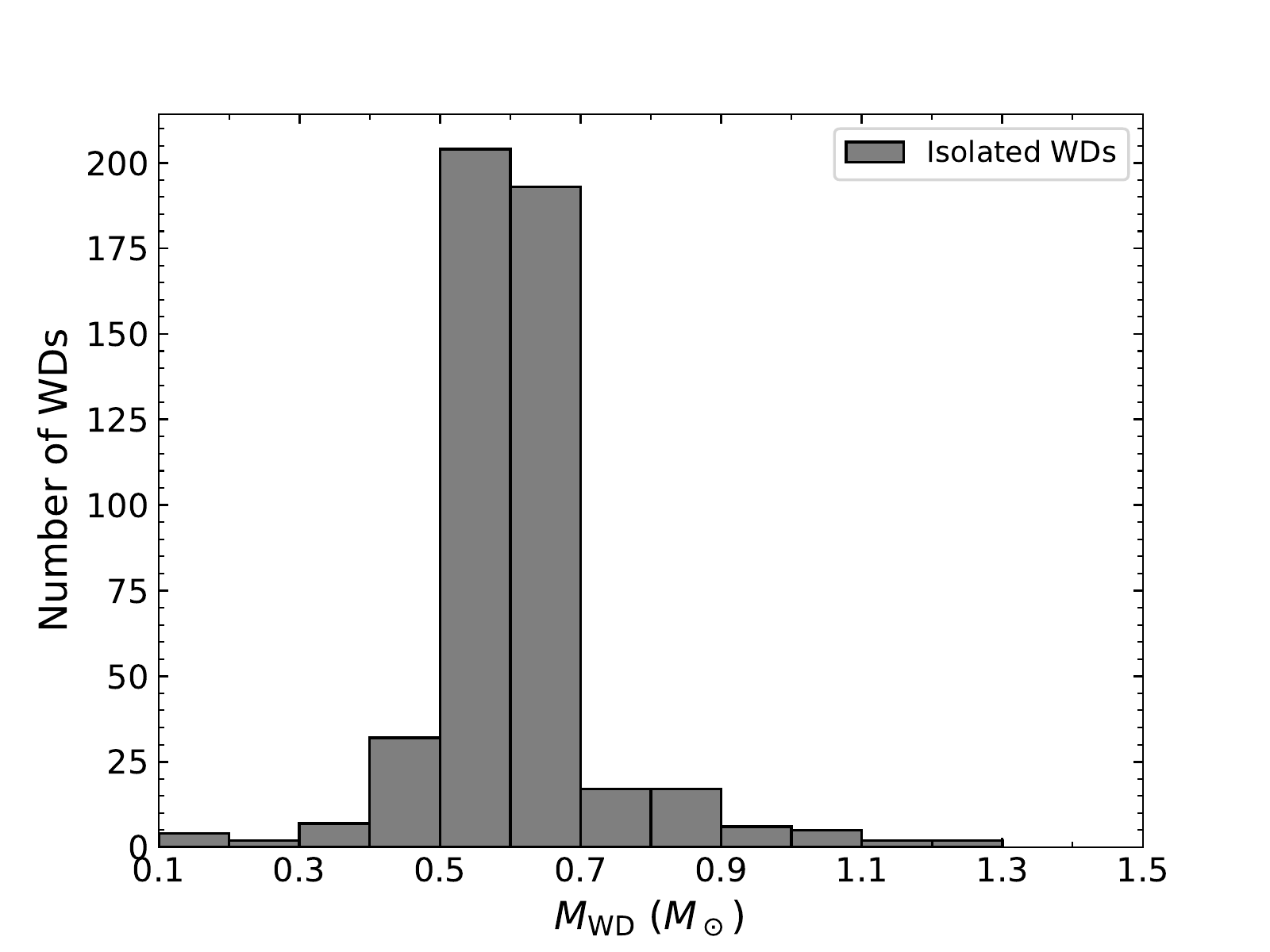}
    \end{subfigure}
    % \vspace{-3mm}
    \begin{subfigure}[b]{0.5\textwidth}
        \includegraphics[width=\textwidth]{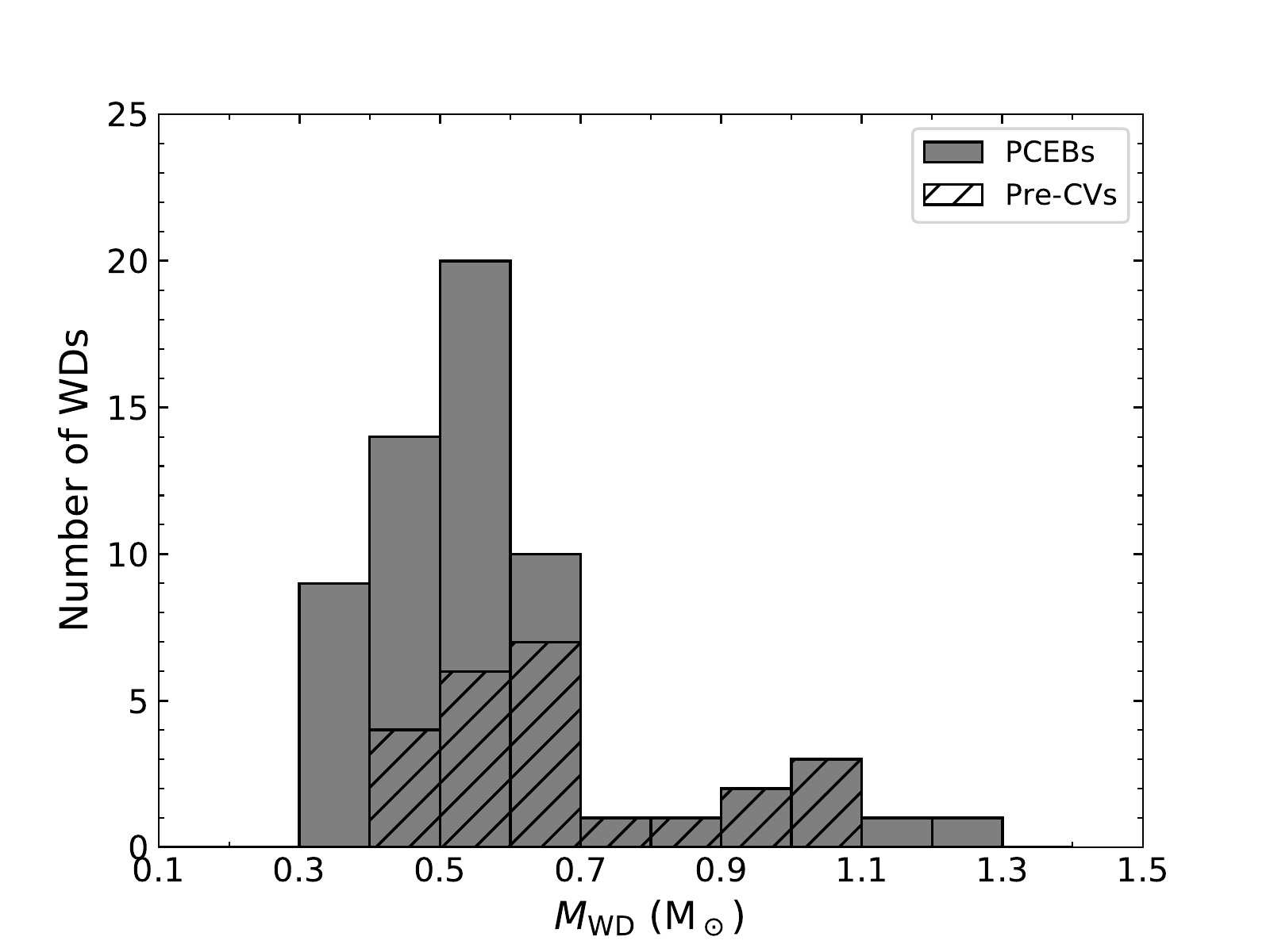}
    \end{subfigure}
    \caption{{\em Upper Panel}: Histograms of WD masses from mCVs \citep[this work plus][hatched]{Suleimanov-2019} and the non-magnetic CVs \citep[grey;][]{Zorotovic-2011}. {\em Middle Panel}: Histogram of masses of isolated WDs from the Sloan Digital Sky Survey \citep{Kepler-2016}. {\em Lower Panel}: Histogram of WD masses from post-common-envelope binaries (grey) and a subset of that sample considered to be representative of progenitors of current CVs \citep[`pre-CVs;'][hatched]{Zorotovic-2011}.}
    \label{fig:hist_comparisons}
\end{figure}

The mass distribution of WDs in the 26 mCVs observed by \nus\ is presented in Fig. \ref{fig:mass_hist}. %To incorporate the uncertainties on the mass into the distribution in Fig. \ref{fig:mass_hist}, we assume each value of \mwd\ results from a normal probability distribution with a mean equal to the measured \mwd\ and standard deviation equal to the $1\sigma$ uncertainty on \mwd. The height of a given histogram bin is now the sum of the probabilities of all the \mwd\ measurements. 
%The mass distribution of the 26 mCVs observed by \nus\ peaks in the range 0.7--0.8 \msun.
The distribution peaks in the range 0.7--0.8 \msun. 
We also present the mCV mass distribution alongside WD mass distributions for different populations of WDs in Fig. \ref{fig:hist_comparisons} in order to draw some comparisons. In the upper panel, we plot the mass distribution of WDs in non-magnetic CVs. To do this we use a sample of CVs that are considered to have `robust' mass measurements \citep[][their table 1]{Zorotovic-2011} and remove the four mCVs\footnote{One of the removed sources is WZ\,Sge, whose status as an mCV remains unclear \citep[see e.g.][]{Matthews-2007}} from that sample, such that only non-magnetic systems remain, for a total of 27 sources. The non-magnetic WDs peak in the range 0.7--0.8 \msun\ and have a weighted average $\bar{M}_{\rm WD}=0.80\pm0.04$ \msun, with $\sigma=0.10$ \msun, consistent with the magnetic WD distribution. %Again, the uncertainty quoted on $\bar{M}_{\rm WD}$ represents the error on the weighted mean%.standard deviation about the weighted mean, while the error on the weighted mean is $0.002$ \msun.

In the middle panel of Fig. \ref{fig:hist_comparisons} we show the mass distribution of isolated WDs. This distribution consists of spectroscopically confirmed WDs from Data Release 12 of the Sloan Digital Sky Survey \citep[SDSS-DR12; see][]{Kepler-2016,Kepler-2016-cat}. We choose hydrogen atmosphere (DA) WDs with a spectral signal-to-noise ratio (S/N) $\geq15$ for a total of 492 WDs with a weighted average $\bar{M}_{\rm WD}=0.53\pm0.03$ \msun and $\sigma=0.15$ \msun. The isolated WD distribution peaks in the range 0.5--0.7 \msun\ and, from Fig. \ref{fig:hist_comparisons}, it is clear just by eye that isolated WDs are preferentially lower mass than both magnetic and non-magnetic WDs in CVs.

The final distribution we compare to, in the lower panel of Fig. \ref{fig:hist_comparisons}, is that of detached post-common-envelope binaries (PCEBs) - the stage in binary evolution that immediately precedes CV formation \citep{Paczynski-1976}. We plot this distribution from the sample presented by \citet{Zorotovic-2011} in their table 2. We highlight in particular a subset of the sample that have CV formation times shorter than the age of the Galaxy and will undergo stable mass-transfer and thus are representative of the present-day CV population, the so-called `pre-CVs' \citep{Zorotovic-2011}. The PCEB distribution peaks in the range 0.5--0.6 \msun\ and the pre-CVs in the range 0.6--0.7 \msun, similar to the isolated WDs, though with a longer tail. The PCEBs and Pre-CVs have a weighted average $\bar{M}_{\rm WD}=0.50\pm0.02$ ($\sigma=0.11$) and $0.53\pm0.04$ ($\sigma=0.12$) \msun, respectively. %The errors on the weighted average for the PCEB and Pre-CV samples are 0.004 and 0.006 \msun, respectively.% ($\sigma_{\rm M}=0.20$ \msun\ for both).

To compare the mCV distribution with other WD distributions we utilise the $k$-sample Anderson-Darling test, where $k=2$ in this instance. We adopt the null hypothesis that two samples are drawn from the same distribution. In comparing the mCVs and non-magnetic CVs \citep{Zorotovic-2011}, we find that we cannot reject the null hypothesis, down to the 10\% level. When comparing the mCV mass distribution with that of the \citet{Kepler-2016} isolated WDs, we find that they must be  drawn from two separate distributions, rejecting the null hypothesis at the $<0.1$\% level. Finally, we find that when we compare the mCV and \citet{Zorotovic-2011} PCEB and Pre-CV samples, we can reject the null hypothesis at the $<0.1$\% level in both instances. Statistically, it appears that mCVs and non-magnetic CVs  exhibit consistent mass distributions, both distinct from those of other types of WD, confirming, and expanding upon, the findings of \citet{Zorotovic-2011}.

\subsection{Comparisons with {\em Swift}/BAT-measured masses}

\begin{figure}
    \centering
    \includegraphics[width=0.5\textwidth]{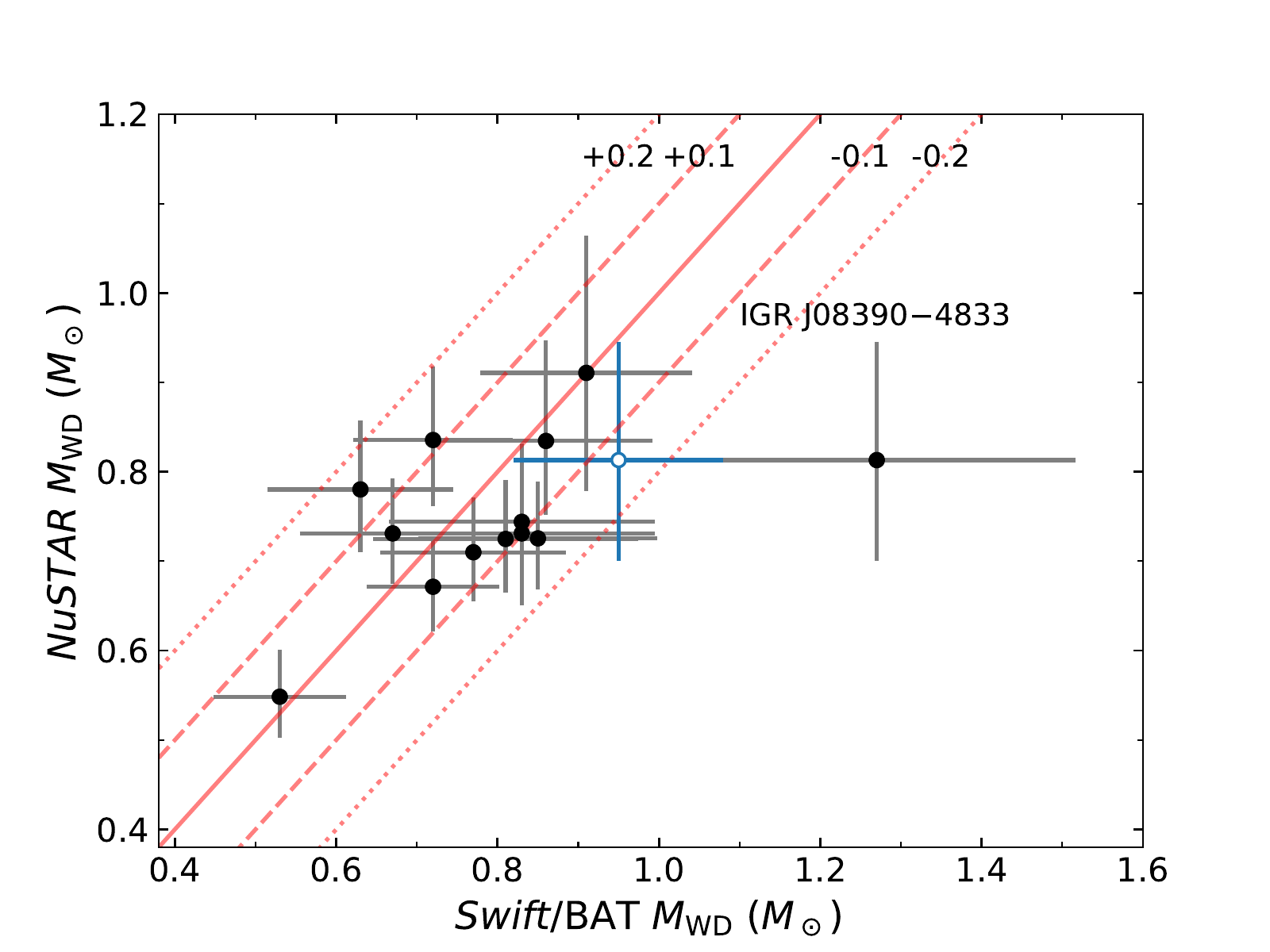}
    \caption{Comparison of WD masses derived by applying the PSR model to \nus\ spectra, with those obtained by applying the same model to {\em Swift}/BAT spectra \citep{Suleimanov-2019}. The solid line is the function \mwd\ (\nus) = \mwd\ (BAT), the dashed lines show \mwd\ (\nus) = \mwd\ (BAT) $\pm0.1$ \msun\ and  the dotted lines show \mwd\ (\nus) = \mwd\ (BAT) $\pm0.2$ \msun. We also plot the value of \mwd\ of IGR\,J08390$-$4833 as measured by {\em INTEGRAL} \citep{Bernardini-2012} with a blue, open circle.}
    \label{fig:Nu_vs_BAT}
\end{figure}

\citet{Suleimanov-2019} analyse a sample of 35 IPs detected in the 70 month {\em Swift}/BAT survey. They use \nus\ to measure the masses for 10 of them (three Legacy targets and 7 previously observed with \nus). Of the remaining 25 IPs in that sample, 13 now have \nus-derived masses. We directly compare the masses we derive from \nus\ spectra with those from {\em Swift}/BAT spectra in Fig. \ref{fig:Nu_vs_BAT}.

The majority of the derived masses are broadly consistent between \nus\ and {\em Swift}/BAT, with a typical scatter $\sim\pm0.1$ \msun. However, Fig. \ref{fig:Nu_vs_BAT} shows one major outlier: IGR\, J08390$-$4833. \citet{Suleimanov-2019} measure \mwd=$1.27\pm0.25$ \msun\ (uncertainty recalculated to 90\% confidence), compared to the $0.81^{+0.13}_{-0.11}$ \msun\ derived from the \nus\ spectra. Previous measurements with {\em INTEGRAL} imply a mass 
%closer to that of 
 consistent with 
the \nus-derived one \citep[\mwd=$0.95\pm0.13$ \msun;][]{Bernardini-2012}. 

IGR\,J08390$-$4833 is located in a complicated region of the sky, where contributions from the Vela supernova remnant (SNR; with which the IP is spatially coincident) and other nearby X-ray sources may cause higher than typical systematic uncertainties for poor angular-resolution measurements, such as %the one
those 
by {\em Swift}-BAT. Indeed, upon closer inspection of the modeled {\em Swift}/BAT spectrum of IGR\,J08390$-$4833, we find that the fit is poor, with a strong excess beyond $30$ keV that can likely be attributed to emission from the SNR and/or other nearby X-ray sources. We therefore suggest that the value in Table \ref{tab:results} is more representative of the true mass of the WD in IGR\,J08390$-$4833, as we were able to isolate and extract photons from the source and background by studying the \nus\ image.

\subsection{Comparisons with masses derived from the iron line complex}

\begin{figure}
    \centering
    \includegraphics[width=0.5\textwidth]{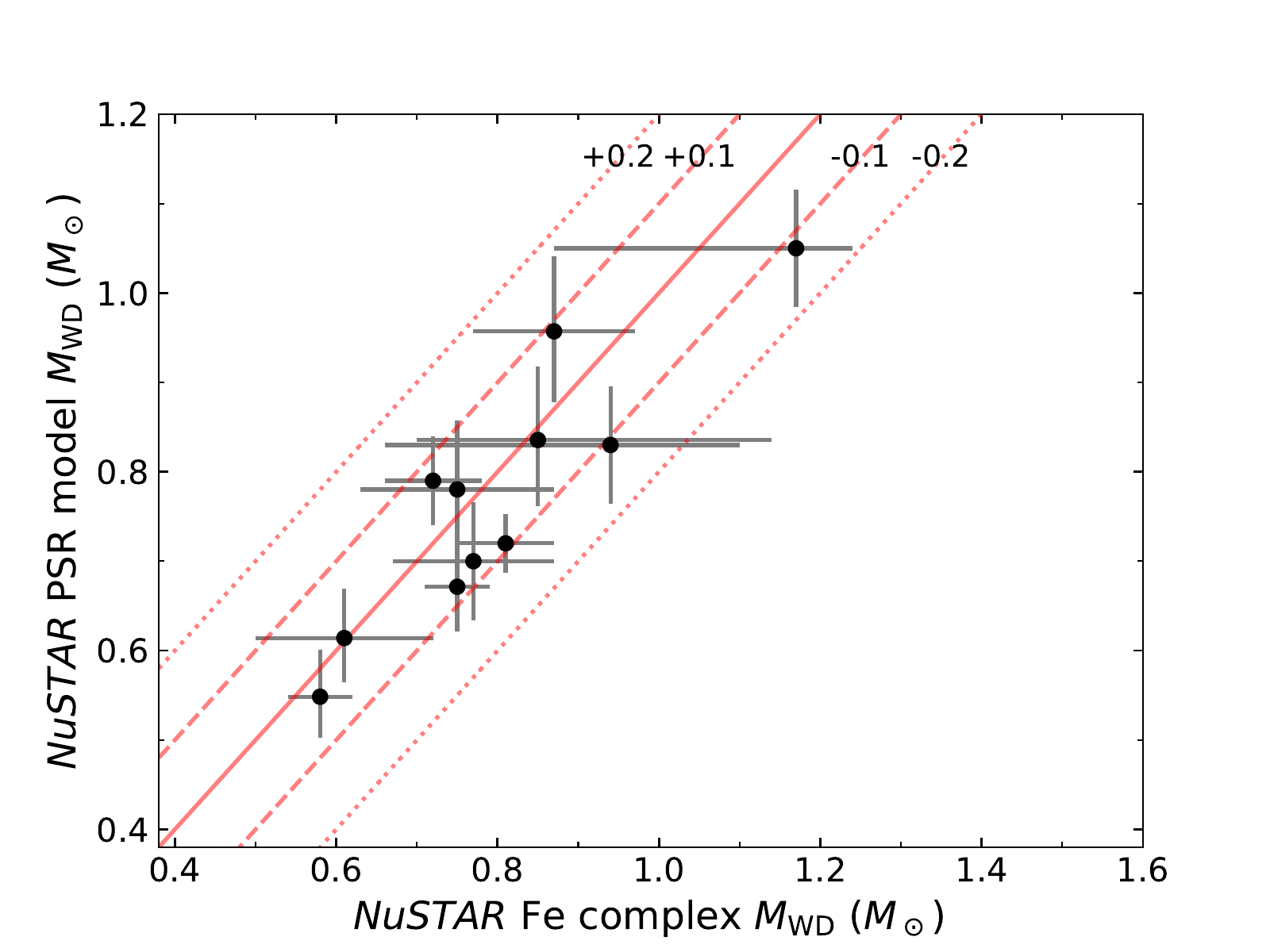}
    \caption{Comparison of WD masses derived from applying the PSR model to the 20--78 keV \nus\ spectra ($y$-axis) with those derived by measuring the intensity ratio of the 7.0 and 6.7 keV Fe lines \citep[][$x$-axis]{Xu-2019}, using the same \nus\ data combined with archival {\em Suzaku} data. The solid line is the function \mwd\ (PSR) = \mwd\ (Fe), the dashed lines show \mwd\ (PSR) = \mwd\ (Fe) $\pm0.1$ \msun\ and  the dotted lines show \mwd\ (PSR) = \mwd\ (Fe) $\pm0.2$ \msun. }
    \label{fig:PSR_vs_Fe}
\end{figure}

\citet{Fujimoto-1997} and \citet{Ezuka-1999} showed that the Fe complex in the $\sim$6--7 keV region of mCV spectra can be used to constrain \mwd. This is achieved by measuring the intensity ratio of the H-like (7.0 keV) and He-like (6.7 keV) components of the Fe complex, which is correlated with the temperature of the PSR. Using the \nus\ observations of mCVs available at the time (some of which are Legacy targets), \citet{Xu-2019} applied a similar methodology to derive \mwd\ for a number of systems. We directly compare the masses derived using the PSR model \citep[][and this work]{Suleimanov-2019} with those derived using the Fe line ratio method on the same \nus\ data \citep[combined with {\em Suzaku} data,][]{Xu-2019} in Fig. \ref{fig:PSR_vs_Fe}.

The PSR model produces results consistent (within 90\% uncertainties) with those of \citet{Xu-2019}, though we note that the PSR model results in smaller uncertainties. The completion of the Legacy survey, which was still in progress at the time of publication of \citet{Xu-2019}, adds 13 more mCVs to the \nus\ archive. In future studies of the Legacy data we will be able to examine if this consistency between the two methodologies holds for all mCVs.

%Xu, Yu and Li use Fe line ratio to measure mass (based on Ezuka and Ishida from what I can tell, but their methodology is lacking - how do they get from I7.0/I6.7 to a mass?).

\subsection{Caveats of the Modeling}

\subsubsection{Choice of Background Extraction Region}

At high energies, the X-ray background becomes more dominant relative to the source photons. In mCVs, where the high energy turnover of the spectrum informs the derived \mwd, it is therefore crucial that the background is measured correctly. We experimented by extracting background spectra for different regions (both on different detector chips and on the same chip as the source) for a subset of targets in the Legacy sample and applying the PSR model. We find that the derived mass remains consistent within uncertainties with the values detailed in Table \ref{tab:results}. 

\subsubsection{Magnetic Field Strength}

The \citet{Suleimanov-2019} PSR model assumes that the dominant cooling mechanism in the PSR is thermal emission (i.e. Brehmsstrahlung). However, if the WD is highly magnetized \citep[$B\gtrsim20$ MG, but note the dependence on \mwd\ and specific accretion rate; see fig. 4 of][]{Wu-1994b} then we might expect cyclotron cooling to compete with thermal emission. In this case, the PSR will be fainter in hard X-rays compared to the model prediction and the resultant mass will be underestimated. 

Magnetic field strength measurements for IPs are rare. However, some of our targets do have optical polarisation measurements that have led to estimates of $B$. Of the IPs in the full \nus\ sample, V405\,Aur, V2400\,Oph, RX\,J2133.7$+$5107 and PQ\,Gem have been suggested to contain WDs with $B\gtrsim20$ MG \citep[see][and references therein]{Ferrario-2015}. The two APs in our target list, V1432\,Aql and BY\,Cam, are very close to being polars and are thus expected to be highly magnetic \citep[$B\gtrsim30$ MG;][and references therein]{Ferrario-2015}. Therefore, the masses of these WDs may be slightly underestimated. Though it is difficult to quantify the mass difference, due to the uncertainty in measurements of $B$ from optical polarimetry, taking cyclotron cooling into account would only push the mass distribution higher. Thus, the conclusion that the mCV mass distribution is distinct from that of PCEBs and isolated WDs remains valid.

\subsubsection{Shock height}
\label{subsec:shockheight}

In our analysis, we have assumed that the shock is sufficiently close to the white dwarf surface such that the difference of the gravitational potential between the surface and the shock can be ignored. In reality, the shock can never be exactly at the surface. Here we investigate the systematic errors this may introduce to our white dwarf mass estimates.

As spectroscopy is relatively unaffected by shock heights $h_{\rm sh}\sim0.1$ \rwd, we instead investigate the shock height in our sample of IPs using their hard X-ray (10--30 keV) spin modulation. At these energies, absorption, the predominant cause of spin modulation below 10 keV, has limited effects. Geometric effects due to tall shocks, on the other hand, can result in a strong hard X-ray spin modulation regardless of photon energy \citep{Mukai-1999}. This effect has been invoked to explain the spin modulation in IPs V709\,Cas \citep{deMartino-2001} and EX\,Hya \citep{Luna-2018}.

The characteristics of spin modulation due to tall shocks are a large amplitude, and modulations that often exhibit flat tops or flat bottoms \citep[see e.g. V709\,Cas;][]{deMartino-2001}. Tall shocks lead to spin modulation because there is a range of viewing angles at which you see the emission from both shocks, so the maximum observed intensity can be twice the minimum. Lower spin modulation amplitude is possible when the footprint of the emission region is large enough that each pole is seen at a range of viewing angles, some allowing visibility of only one pole, others for both poles to be observable simultaneously. Small amplitude modulations require that poles largely remain in the one-pole only viewing zone with at most small, partial excursions into the two-pole viewing zone, or vice versa. This is likely to result in flat-bottomed (flat-topped) light curves.

We have examined the spin-folded 10--30 keV light curves of our targets. The APs V1432\,Aql and BY\,Cam have large-amplitude modulations, as expected, since they are presumed to accrete onto one pole at a time \citep{Staubert-2003,Pavlenko-2013}. The IPs in our sample have low-amplitude modulations and do not collectively fit our expectations for tall shock systems. They all appear to show small but statistically significant spin modulations, of order $\pm$10\% of the mean. The spin modulations are sometimes single peaked, sometimes double-peaked, and sometimes complex. Since spectral fits below 10 keV often indicate the presence of absorber components with hydrogen column density $N_{\rm H}$ up to several times 10$^{23}$ cm$^{-2}$, or Compton optical depths of a few tenths of unity, we believe that the observed level of hard X-ray absorption can be expected due to variable complex absorption.

Diametrically opposite X-ray emission regions 0.1 \rwd\ above the surface are observable for viewing angles 65--115$^\circ$. We argue that this is not generally the case for IPs in our sample, as it would frequently lead to obvious symptoms of a tall shock, as has been observed in V709\,Cas \citep{Mukai-2015}. The range is 82--98$^\circ$ for emission regions 0.01 \rwd\ above the surface. This range of viewing angle is sufficently small compared to the expected angular extent of the emission region that it is plausible for the resulting hard X-ray modulation to be smooth (i.e., not flat-topped or flat-bottomed) and small in amplitude, as we argue. This line of reasoning suggests that the systematic uncertainties due to tall shocks for our sample, as a group, is of order a few percent.

We can extrapolate the case of EX\,Hya to place rough estimates on the shock heights of some of the mCVs in our sample and test the above discussion. \citet{Luna-2018} 
%place an upper limit on the 
estimate a 
shock height $h_{\rm sh}\gtrsim0.9$ \rwd\ for EX\,Hya, which  exhibits a luminosity $L\sim8\times10^{31}$ erg s$^{-1}$ \citep{Suleimanov-2019}.\footnote{The inequality sign reported by \citet{Luna-2018} is incorrect according to their fig. 4, we use the correct one ($\gtrsim$) here.} \footnote{An alternative assessment of the shock height of EX\,Hya by \citet{Hayashi-2014}, using a detailed X-ray spectral model of the post-shock accretion column, gives a shock height of 0.33 \rwd. We therefore choose $h_{\rm sh}=0.9$ \rwd\ as a fiducial value for EX\,Hya for this order of magnitude estimate.}
Shock height $h_{\rm sh}$ is inversely proportional to local mass accretion rate \citep{Mukai-1999}. If we assume that the footprint area of the shock above the surface of the WD is the same for all mCVs, then $h_{\rm sh}$ varies inversely with overall mass accretion rate (\mdot) and therefore luminosity (along with a mass dependence). Based on this luminosity dependence, the faintest mCV in our target list, BY\,Cam with a luminosity $L=1.3\times10^{33}$ erg s$^{-1}$, %would therefore 
can be estimated to 
have $h_{\rm sh}\sim0.06$ \rwd, and brighter mCVs should have shorter shocks. The assumption that the footprint area is constant between sources is not completely secure, it is unclear how they vary amongst mCVs with different values of $B$ and \pspin\ \citep[for example see e.g.][]{Scaringi-2010}. Nevertheless, we may use the above extrapolation as an order of magnitude estimate of $h_{\rm sh}$, showing that systematic uncertainties in mass  
due to tall shocks should 
%are indeed 
typically  be of order a few percent.

This does not preclude the possibility of a more significant systematic error for individual objects. Among the Legacy sample, the hard X-ray spin modulation amplitude is of order $\pm$20\% or greater for IGR\,J16547$-$1916, AO\,Psc, V405\,Aur and FO\,Aqr. These are the IPs for which larger systematic errors are most likely.
 
The conclusion that WDs in IPs are more massive than in the field remains secure in any case, since tall shock effects can only lead to underestimates of \mwd. %This effect does affect our ability to determine whether WDs in IPs are less massive than the WDs in non-magnetic CVs.

\subsection{Selection Effects}
\label{subsec:sel_effects}

With any survey of a specific class of astrophysical objects, one must consider the possible biases that may arise from the way the sample is selected. We discuss potential selection effects of our sample below and any subsequent effects they may have on our results.

\subsubsection{Source flux}
When devising the \nus\ Legacy Survey of mCVs, we chose our sample based on the flux in the {\em Swift}/BAT 70 month catalogue \citep{Baumgartner-2013,Mukai-2017}. We chose the 25 brightest mCVs (in the BAT energy band; 14--195 keV) that had not previously been observed by \nus, such that the limiting flux of our sample is $\sim1.7\times10^{-11}$ erg cm$^{-2}$ s$^{-1}$. Of the 25 initial targets, two were not observed (IGR\,J16500-3307 and IGR\,J04571+4527), two were not detected by \nus\ (DO\,Dra and IGR\,J14536-5522) and two did not provide high enough S/N spectra to accurately constrain the mass using the method described in Section \ref{sec:obs} (XY\,Ari and RX\,J2015.6+3711). 

It could be assumed that, considering the sample is flux selected in the hard-X-ray band, the results may be biased towards higher masses. Though it is impossible to remove all potential bias arising from a flux-limited sample, our target selection seeks to reduce bias towards higher masses as much as possible. \citet{Suleimanov-2019} measure masses for 35 objects from the {\em Swift}/BAT 70 month catalogue, which is the majority of the confirmed IPs in the catalogue. Their limiting flux is approximately half of ours \citep[V1033\,Cas; $8.43\times10^{-12}$ erg cm$^{-2}$ s$^{-1}$;][]{Mukai-2017}. Of those 35 IPs, there are now \nus-measured masses for 23 of them, and we find that they generally agree with the {\em Swift}/BAT-measured masses but with smaller uncertainties \citep[Fig. \ref{fig:Nu_vs_BAT}; also fig. 8 of][]{Suleimanov-2019}. We can therefore reasonably assume that the remaining sources below our flux threshold have accurate {\em Swift}/BAT-measured masses, and these remaining sources are not biased toward any mass, high or low. In addition, there are only {\em five} confirmed IPs that are not detected by {\em Swift}/BAT (HT\,Cam, DW\,Cnc, UU\,Col, V1323\,Her and WX\,Pyx). The limited X-ray information available regarding these objects suggests that they exhibit a range of shock temperatures \citep[e.g.][]{Schlegel-2005,deMartino-2006b,Nucita-2019} and therefore likely a range of masses. Any mass bias that exists due to the way in which we selected our sample is unlikely to be large. 

However, we must make it clear that our sample selection does not preclude the existence of a population of (possibly low-mass) mCVs that may not have been identified as such due to their non-detection by X-ray observatories. This cannot be mitigated with statistical analysis and we base our results and conclusions on the known, visible population of mCVs.

%Firstly, we find that the 14--195 keV {\em Swift}/BAT flux does not have a strong dependence on the 2--10 keV flux.\footnote{\href{https://asd.gsfc.nasa.gov/Koji.Mukai/iphome/catalog/omegaomega.html}{https://asd.gsfc.nasa.gov/Koji.Mukai/iphome/catalog/omegaomega.html}} 

% In a study of the space-density of IPs, \citet{Pretorius-2014} choose a flux cut of $>2.5\times10^{-11}$ erg cm$^{-2}$ s$^{-1}$ and a Galactic latitude cut of $\lvert b\rvert > 5^{\circ}$ to infer $1^{+1}_{-0.5}\times10^{-7}$ pc$^{-3}$. Barring a population of unobserved faint IPs, the \citet{Pretorius-2014} sample may be  considered a reasonable representation of the intrinsic IP population. The Legacy (plus \citealt{Suleimanov-2019}) sample not only considers lower hard X-ray fluxes, but makes no cut on source location (see Fig. \ref{fig:CV_map}). {\em Gaia}-derived distances to our targets range from $d=56.8\pm0.1$pc to $d=56.8\pm0.1$ to $d=2.1^{+0.3}_{-0.2}$ kpc \citep{Bailer-Jones-2018}. We therefore propose that the sources discussed in this work (and therefore their derived mass distribution) are representative of the intrinsic mCV population. 

\subsubsection{Origin of the target's X-ray discovery}

Many of the X-ray observatories that discovered the mCVs in this work operate at hard X-ray energies. For example, all of the `IGR' labelled IPs in our sample were discovered by the IBIS instrument onboard {\em INTEGRAL}, which operates at energies $>15$ keV. Considering the fraction of the total flux emitted in the hard band increases with \mwd, hard X-ray instruments are more likely to detect massive WDs. Though a number of sources in the full \nus\ sample were first detected in X-rays by {\em ROSAT} \citep[0.1--2 keV;][]{Truemper-1982}, the majority were discovered by instruments with some hard X-ray sensitivity (e.g. {\em Ariel V}; 1.5--20 keV, {\em Uhuru}; 2--20 keV and {\em HEAO-1}; 0.25--25 keV). We cannot discount the possibility of a bias towards higher masses. Though we note here that the five IGR sources in the full \nus\ sample %\citep[those in Table \ref{tab:results} and V2731\,Oph;][]{Suleimanov-2019} 
in addition to V667\,Pup which was discovered by {\em Swift}/BAT ($>15$ keV), range from \mwd=0.69--1.06 \msun, similar to the 8 {\em ROSAT}-discovered sources (0.61--1.05 \msun).

\subsection{The CV mass problem}

Considering the discussion above, we have shown that mCVs, like their non-magnetic counterparts, are preferentially more massive than both isolated WDs and PCEBs, consistent with previous surveys with non-imaging hard X-ray telescopes \citep[e.g.][]{Yuasa-2010,Bernardini-2012,Suleimanov-2019}. Whilst we cannot dismiss the possibility that unidentified systematic uncertainties in the mass measurements of both non-magnetic and magnetic systems contribute to this observed difference, we can only discuss the origin of the discrepancy in the context of the existing observations. Therefore, how do we reconcile this with theoretical predictions? The classic picture of CV formation starts with a wide main-sequence -- main-sequence (MS -- MS) binary, whereby one of the binary components becomes a red giant and fills its Roche lobe, initiating mass transfer on to the companion. The unstable nature of this mass transfer leads to a common envelope (CE) phase and the orbital separation is reduced through drag forces within the envelope. Once the envelope is expelled, what is left behind is a close (yet detached) WD -- MS binary, i.e., the PCEB scenario discussed above. Upon further reduction of the binary separation \citep[through angular momentum loss by a combination of gravitational radiation and magnetic braking;][]{Knigge-2011}, the binary will then initiate the second mass transfer stage that defines CVs \citep{Paczynski-1976}. A consequence of this evolutionary path is that WDs in CVs should have a (slightly) lower mean mass than that of isolated WDs \citep[][]{Politano-1996}.

We now know that observationally, this is not true. \citet{Zorotovic-2011} show that $\bar{M}_{\rm WD}=0.83$ \msun\ from a non-magnetic CV sample free of observational biases related to WD mass. We show from spectral modeling of high quality \nus\ observations that the weighted average mass of WDs in mCVs is similarly high, $\bar{M}_{\rm WD}=0.77\pm0.02$ \msun, with $\sigma=0.10$ \msun. Matters are complicated further by the fact that studies of PCEBs, i.e. the precursor to CVs, have revealed that the observed WD mass distribution of these objects is in good agreement with theoretical predictions \citep[e.g.][]{Zorotovic-2011,Toonen-2013,Camacho-2014}, meaning that the problem is not due to an underlying misunderstanding of CE evolution.

Mass-growth does not appear to solve the problem, nor does a short phase of thermal timescale mass transfer, at least for non-magnetic CVs \citep{Wijnen-2015}. It has long been suggested that nova explosions should prevent mass-growth from occurring, if the amount of mass expelled in the explosion is more than the amount accreted between outbursts, as is predicted by a number of theoretical models (\citealt{Prialnik-1995,Yaron-2005}; but see below). \citet{Schreiber-2016} suggest that consequential angular momentum loss (CAML) may solve the WD mass problem in non-magnetic CVs. The CAML hypothesis suggests that angular momentum loss driven by mass transfer (e.g. frictional angular momentum loss through nova explosions) is more effective in lower mass systems, resulting in mass transfer becoming unstable in such systems. CVs therefore have preferentially higher mass WDs. Our results may indicate that CAML works similarly for both magnetic and non-magnetic CVs.% though nova explosions are less common in mCVs (GK\,Per is the prototypical IP nova).

Another potential solution to the CV mass problem could be that the amount of mass expelled by a nova explosion is less than some theoretical models predict. %, resulting in a net mass gain over an accretion--nova cycle. 
According to a number of simulations \citep[e.g.][]{Hillman-2015}, there is a region of the parameter space where \mwd\ grows after successive accretion--nova cycles. While this region was limited to very high mass WDs in most models, hence did not address the observational discrepancy, more recent hydrodynamical simulations of classical novae by \citet{Starrfield-2020} suggest that WDs with masses in the range 0.6--1.35 \msun\ can grow in mass through accretion--nova cycles. The fact that nova models are seen to contradict one another on the topic of mass-growth, shows that there is no clear consensus on the matter.

% Alternatively, \citet{Suleimanov-2019} suggest that IPs could be in a low mass accretion rate (\mdot) state and may exhibit short, rare and barely visible outbursts akin to dwarf novae (again, see e.g. GK\,Per). During a dwarf nova outburst, the {\em local} \mdot\ on the surface of the WD will be much higher in mCVs than for regular CVs due to matter only accreting onto the magnetic poles. An open question is therefore posed: could the local \mdot\ be high enough during a dwarf nova outburst that stable nuclear burning of hydrogen occurs? In such a scenario, one might expect an increase in \mwd\ \citep[e.g.][]{Shen-2007}. The fact that some IPs do exhibit nova explosions, however, suggests that stable burning might not be the solution.

%\citet{Suleimanov-2019} find that most IPs show lower mass-accretion rates than what is expected for regular CVs \citep{Howell-2001} and suggest that IPs are therefore in a low mass accretion rate state, showing short, rare, barely visible nova outbursts. This suggests that mCVs could grow in mass through accretion, and the short nova eruptions do not remove as much mass as models suggest. 

\section{Conclusions}
\label{sec:conclusions}

We have conducted the first dedicated survey of mCVs with an imaging hard X-ray telescope in order to derive the mass distribution of magnetic WDs. Adding the results of this survey to those of the 7 IPs previously observed by \nus\ \citep{Suleimanov-2019} brings the total number of accreting magnetic WD masses constrained with \nus\ to 26. This is the largest single sample of mCV masses constrained with imaging telescopes to date.

We utilized the PSR X-ray spectral model \citep{Suleimanov-2016,Suleimanov-2019} to derive \mwd. For FO\,Aqr, we 
confirmed the \citet{Suleimanov-2019} 
%were also able to obtain an 
estimate of \rmag\ based on the measurement of a break in the aperiodic power spectrum. We find the weighted average of all 26 mCVs to be $\bar{M}_{\rm WD}=0.77\pm0.02$ \msun\ with a standard deviation of 0.10 \msun. Statistically, the mass distribution is consistent with that of WDs in non-magnetic CVs, i.e. accreting WDs, whether magnetic or not, appear to preferentially have higher masses than both isolated WDs and the precursors to CVs, PCEBs. This compounds the CV mass problem, i.e. the discrepancy between observations and theory surrounding masses of accreting WDs. We speculate that consequential angular momentum loss \citep{Schreiber-2016} may play a role in this discrepancy, but also note that our understanding of how \mwd\ changes over accretion--nova cycles may also be incomplete. %We speculate that...

\subsection{Future Work}

The Legacy dataset that resulted in this work is extensive, and PSR modeling of the $>20$ keV spectra is just one of the analysis approaches we can take. \citet{Xu-2019} showed, with a small number of \nus\ spectra, that there is a wealth of information embedded within the Fe line complex that can lead to an independent derivation of PSR temperature and therefore mass. In addition a study of the full 3--78 keV spectra, will allow us to conduct an in-depth analysis of reflection and partial covering in mCVs, allowing estimates of the shock height, as well as an alternative spectral fitting method to measure \mwd. %and if and how such processes may affect the derived \mwd.

\section*{Data Availability Statement}

The data underlying this article are publicly available in the {\em NuSTAR} Master Catalog (NUMASTER) at \href{https://heasarc.gsfc.nasa.gov/W3Browse/nustar/numaster.html}{https://heasarc.gsfc.nasa.gov/W3Browse/nustar/numaster.html}. The Observation ID for each target is listed in Table \ref{tab:observations}.% Reduced spectra are available upon request. %Hmmm not sure about this, I'll think about it.

\section*{Acknowledgements}

We would like to thank the anonymous referee for useful comments which helped improve the manuscript. We also thank Fiona Harrison and the {\em NuSTAR} team for approving the mCV Legacy programme. AWS would like to thank Rich Plotkin for useful discussions regarding some of the results in this work. AWS would also like to thank Phil Uttley, Abigail Stevens and Peter Bult for advice regarding timing analysis. COH acknowledges support from NSERC Discovery Grant RGPIN-2016-04602, and a Discovery Accelerator Supplement. VD and VFS thank the Russian Science Foundation (grant 19-12-00423) for financial support. VFS also thanks Deutsche  Forschungsgemeinschaft (DFG) for financial support (grant WE 1312/51-1). DJKB acknowledges support from the Royal Society. BWG acknowledges support under NASA contract No. NNG08FD60C. JH acknowledges support from an appointment to the NASA Postdoctoral Program at the Goddard Space Flight Center, administered by the USRA through a contract with NASA. JJ acknowledges support by the Tsinghua Shui'mu Fellowship and the Tsinghua Astrophysics Outstanding Fellowship. RML acknowledges the support of NASA through Hubble Fellowship Program grant HST-HF2-51440.001. GRS acknowledges support from an NSERC Discovery Grant (RGPIN-2016-06569).

This work made use of data from the {\em NuSTAR} mission, a project led by the California Institute of Technology, managed by the Jet Propulsion Laboratory, and funded by the National Aeronautics and Space Administration. We thank the {\it NuSTAR} Operations, Software and  Calibration teams for support with the execution and analysis of these observations.  This research has made use of the {\it NuSTAR}  Data Analysis Software (NuSTARDAS) jointly developed by the ASI Science Data Center (ASDC, Italy) and the California Institute of Technology (USA). This research has made use of data and/or software provided by the High Energy Astrophysics Science Archive Research Center (HEASARC), which is a service of the Astrophysics Science Division at NASA/GSFC.

%%%%%%%%%%%%%%%%%%%%%%%%%%%%%%%%%%%%%%%%%%%%%%%%%%



\bibliographystyle{mnras}
\bibliography{NuSTAR_Legacy.ACCEPTED.arxiv.bib}

%%%%%%%%%%%%%%%%%%%%%%%%%%%%%%%%%%%%%%%%%%%%%%%%%%

%%%%%%%%%%%%%%%%% APPENDICES %%%%%%%%%%%%%%%%%%%%%

\appendix

\section{Spectral figures}

We plot the \nus\ spectra of the Legacy survey targets (excluding FO\,Aqr; see Fig. \ref{fig:FO_Aqr} in Fig. \ref{fig:spec_figs}.

\begin{figure*}
    \centering
    \begin{subfigure}[b]{0.31\textwidth}
        \includegraphics[width=\textwidth]{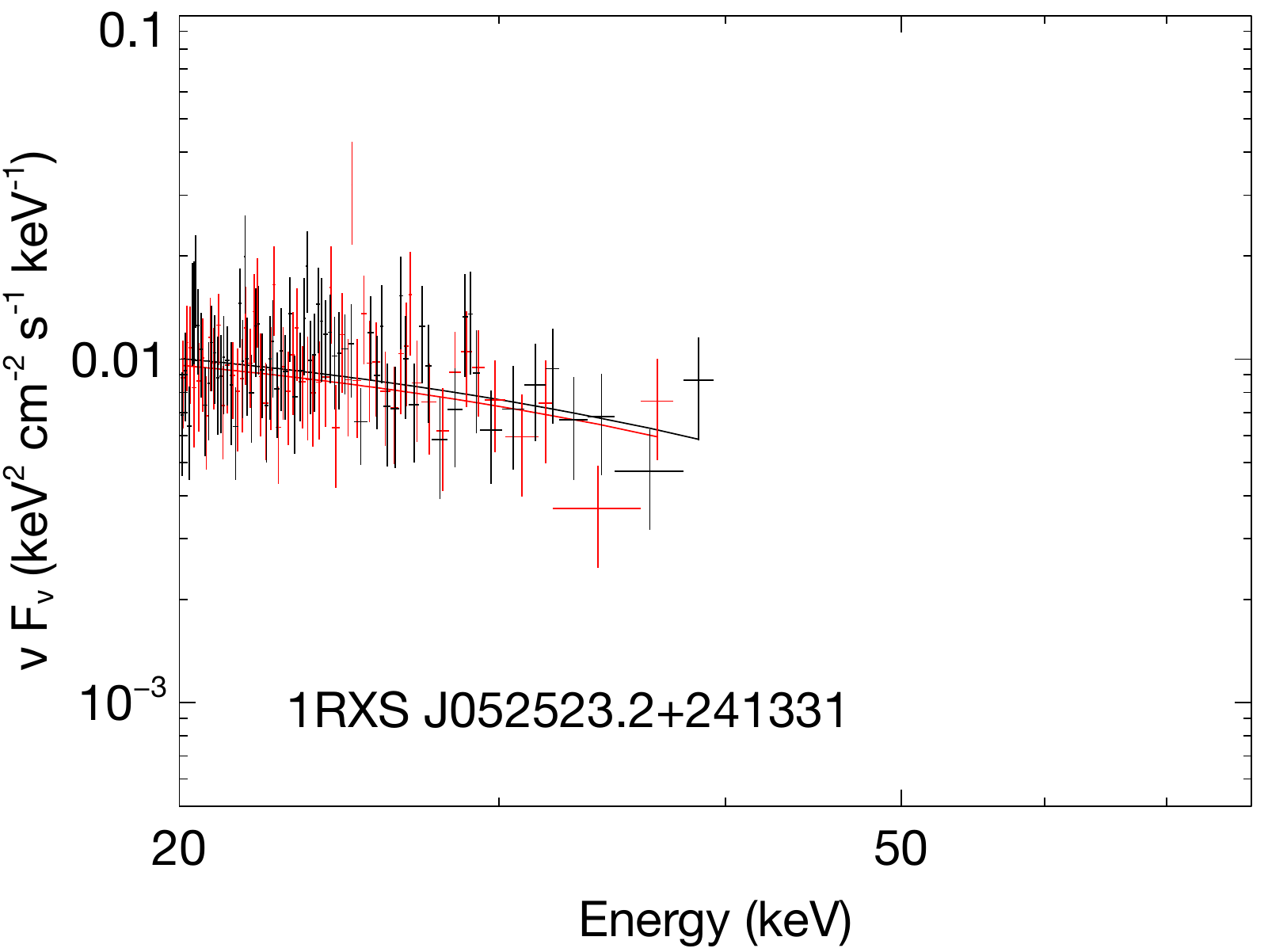}
    % \vspace{-5mm}
    \end{subfigure}
    % \vspace{-3mm}
    \begin{subfigure}[b]{0.31\textwidth}
        \includegraphics[width=\textwidth]{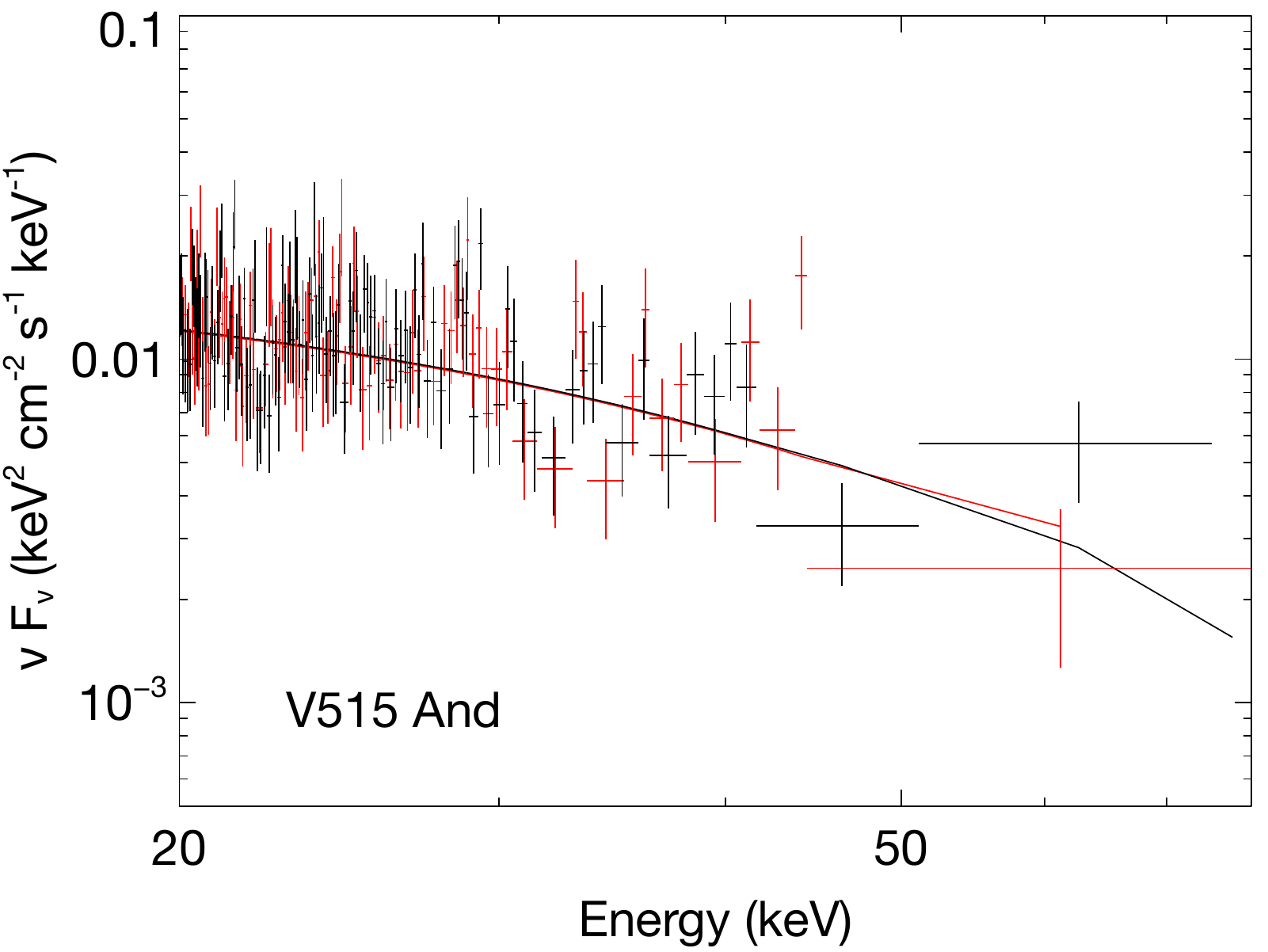}
    \end{subfigure}
    % \vspace{-3mm}
    \begin{subfigure}[b]{0.31\textwidth}
        \includegraphics[width=\textwidth]{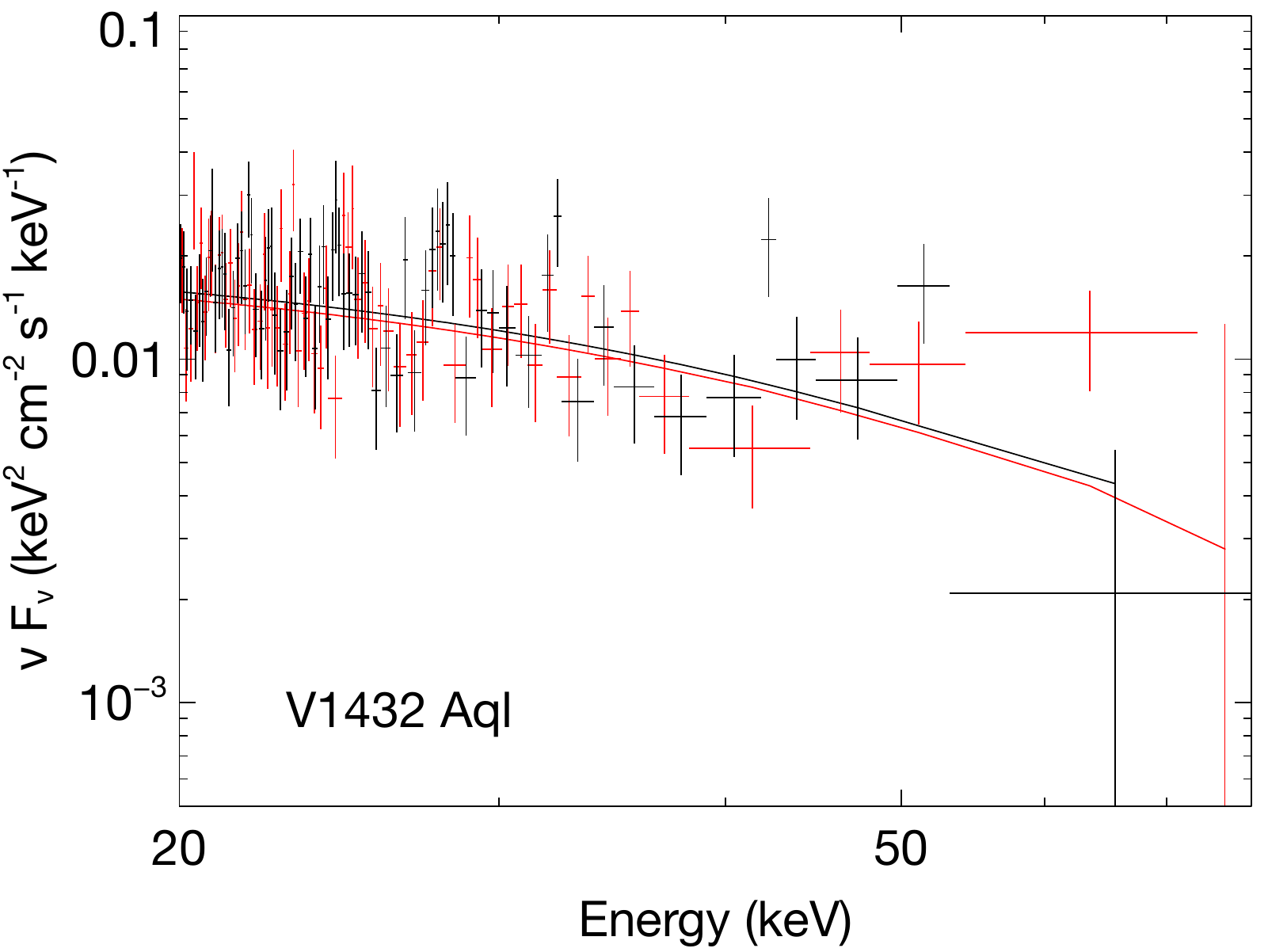}
    \end{subfigure}
    \begin{subfigure}[b]{0.31\textwidth}
        \includegraphics[width=\textwidth]{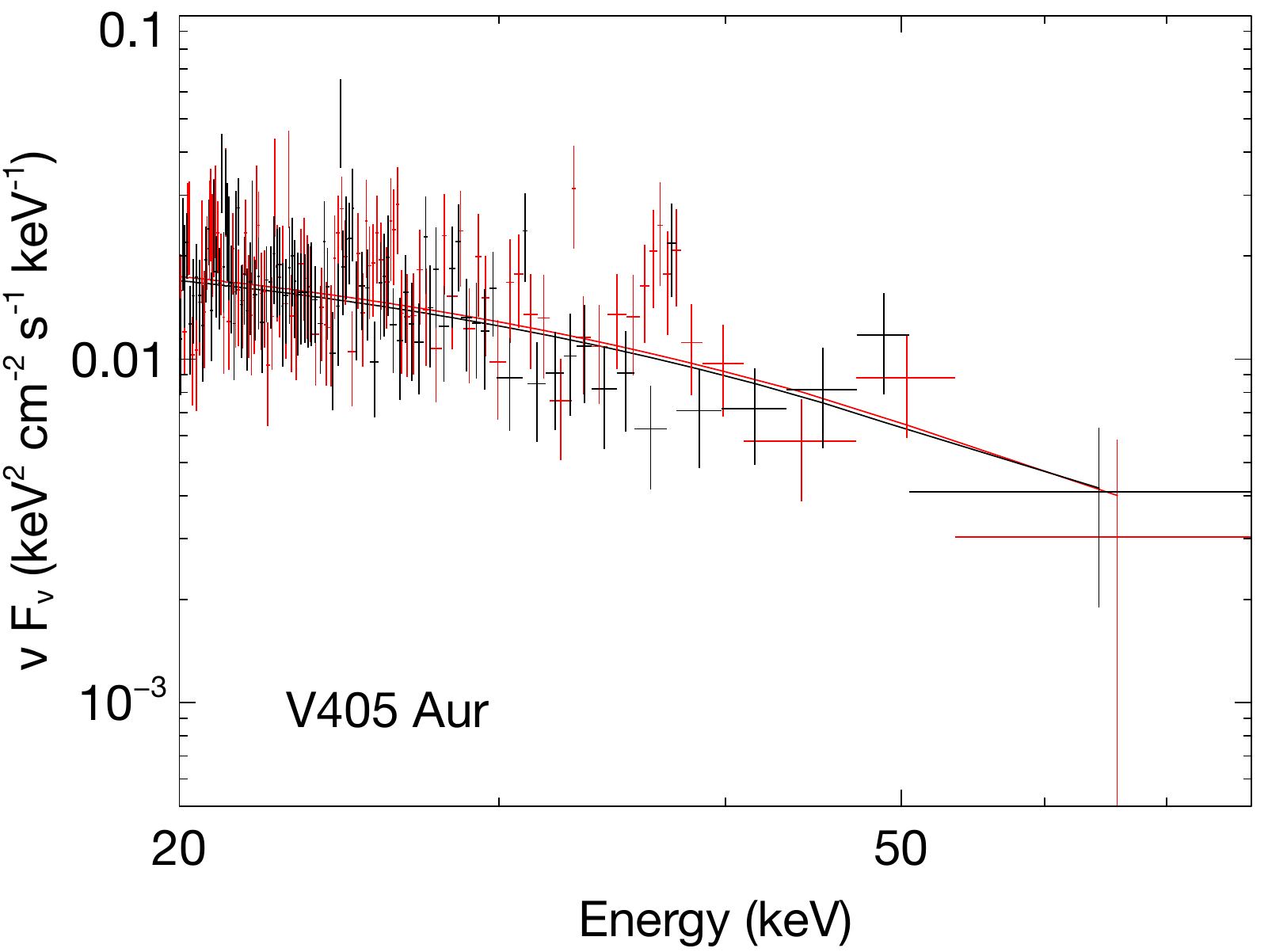}
    \end{subfigure}
    \begin{subfigure}[b]{0.31\textwidth}
        \includegraphics[width=\textwidth]{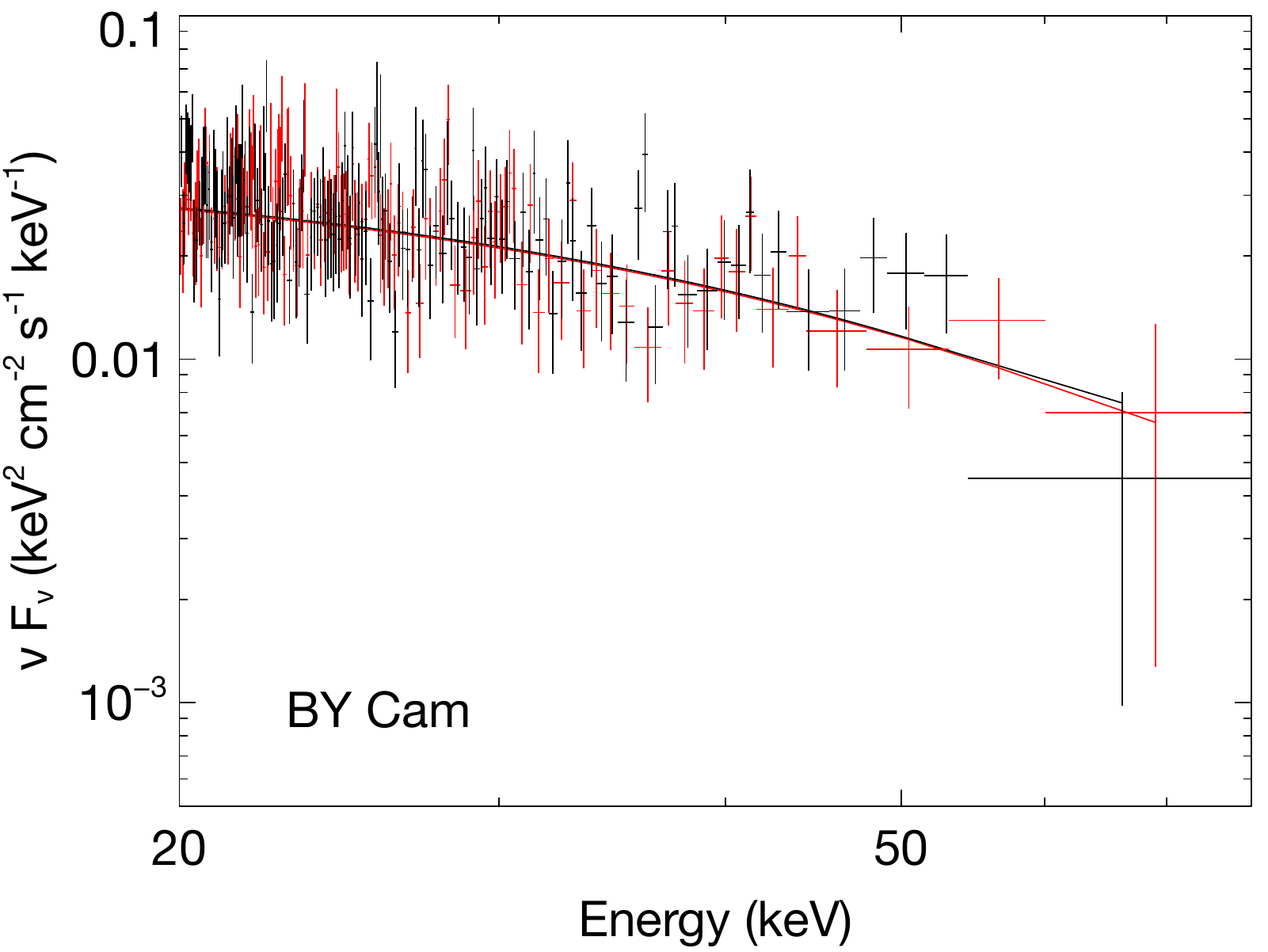}
    \end{subfigure}
    \begin{subfigure}[b]{0.31\textwidth}
        \includegraphics[width=\textwidth]{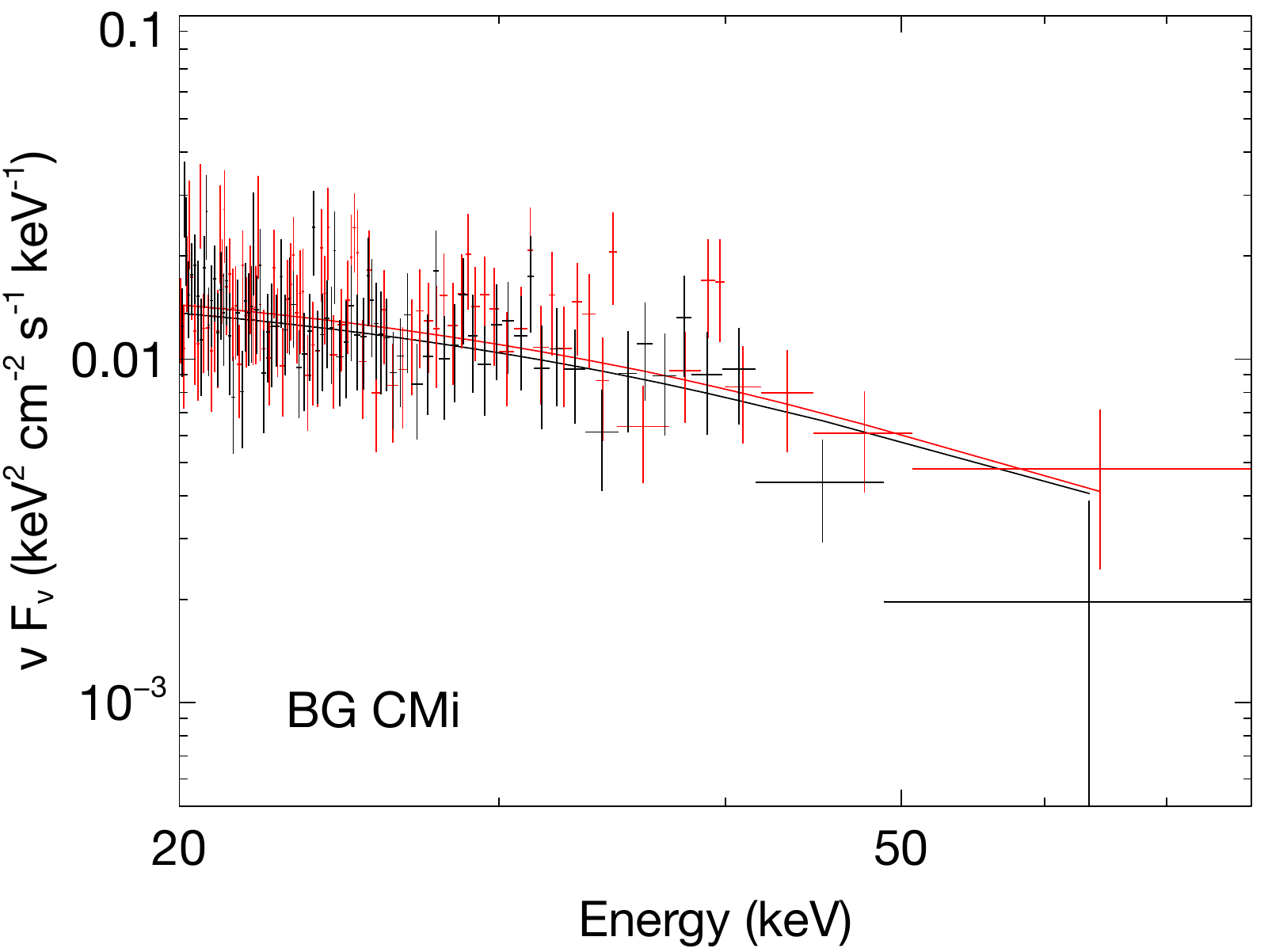}
    \end{subfigure}    
    \begin{subfigure}[b]{0.31\textwidth}
        \includegraphics[width=\textwidth]{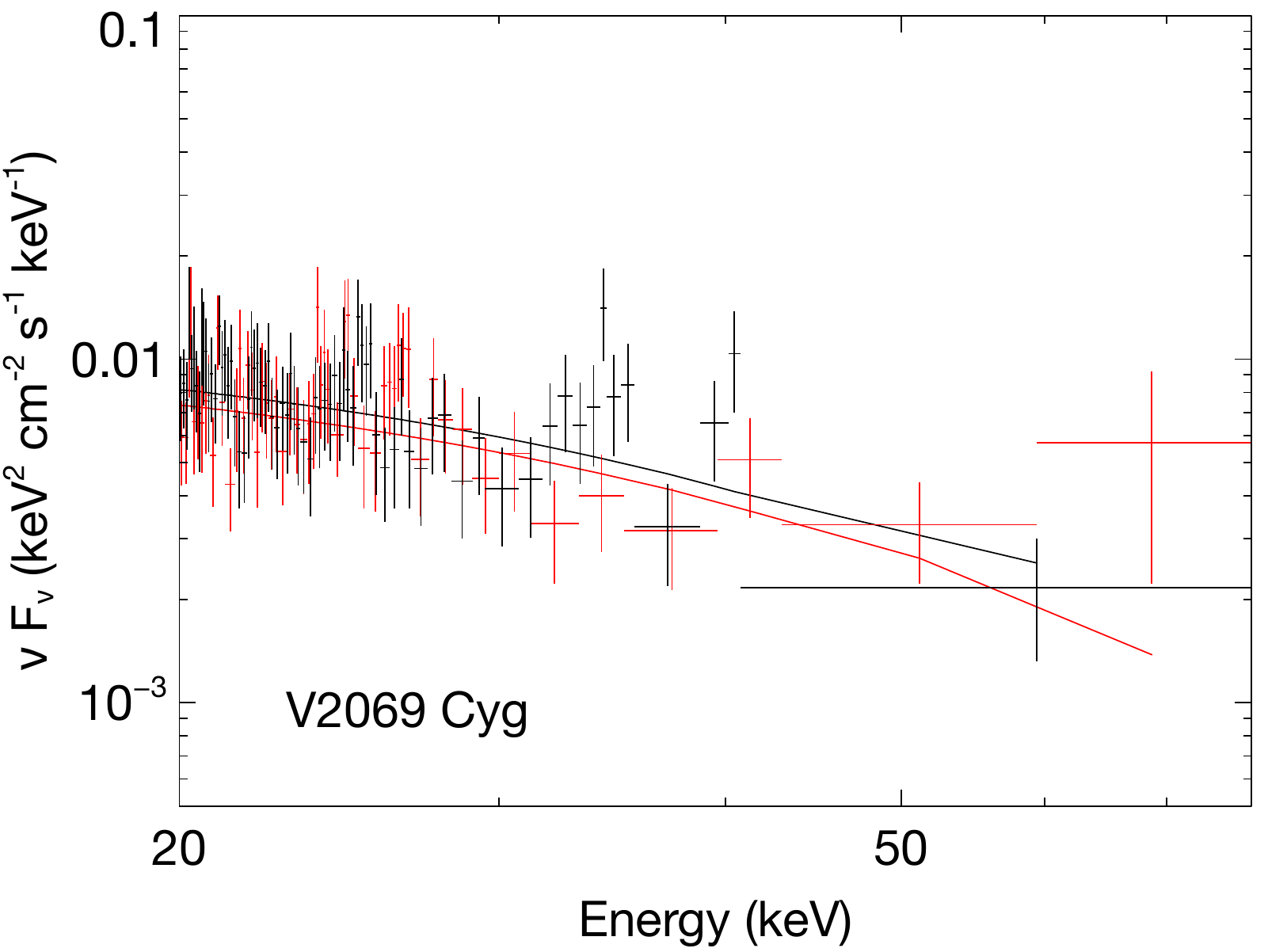}
    \end{subfigure}
    \begin{subfigure}[b]{0.31\textwidth}
        \includegraphics[width=\textwidth]{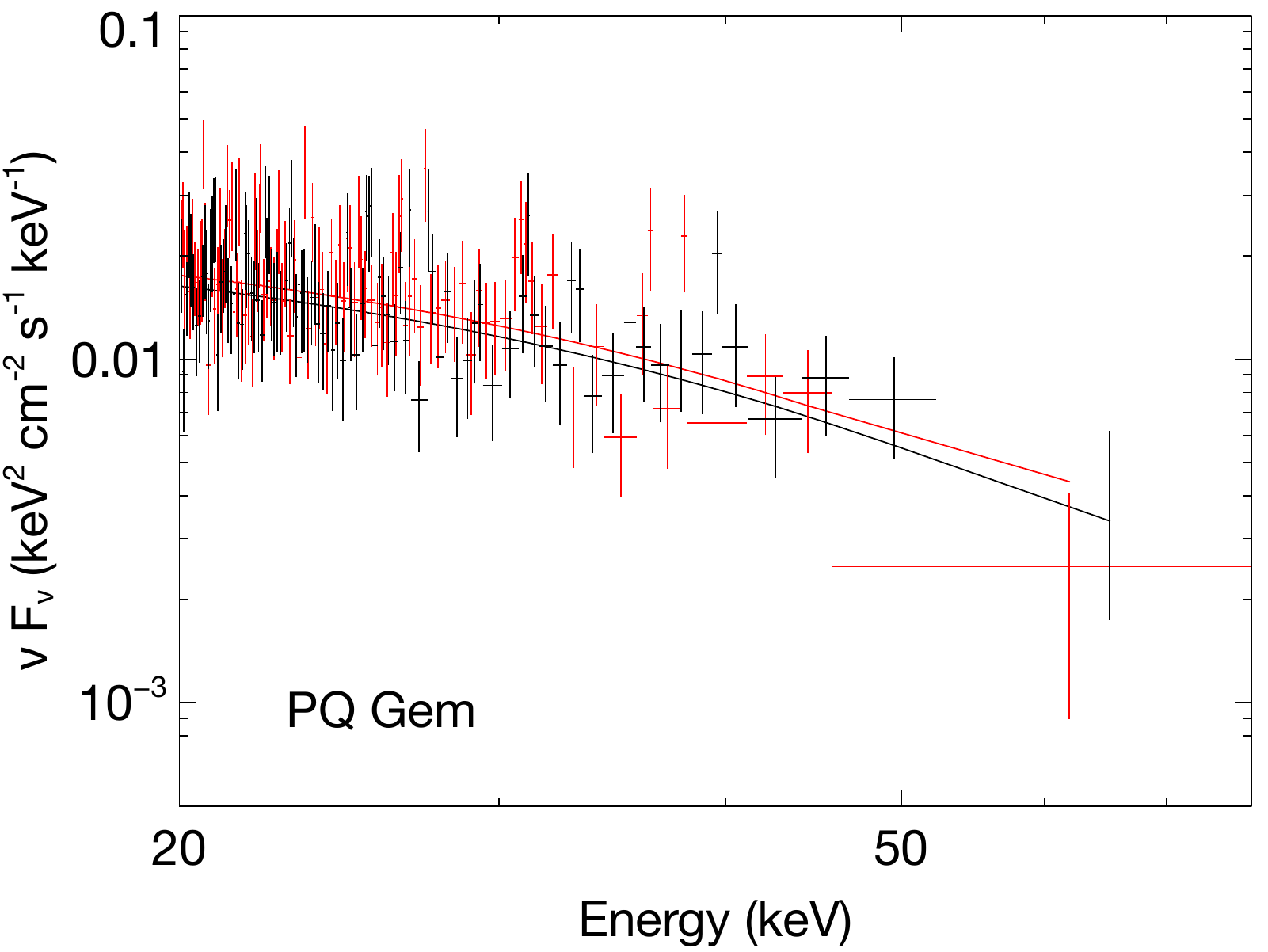}
    \end{subfigure}
    \begin{subfigure}[b]{0.31\textwidth}
        \includegraphics[width=\textwidth]{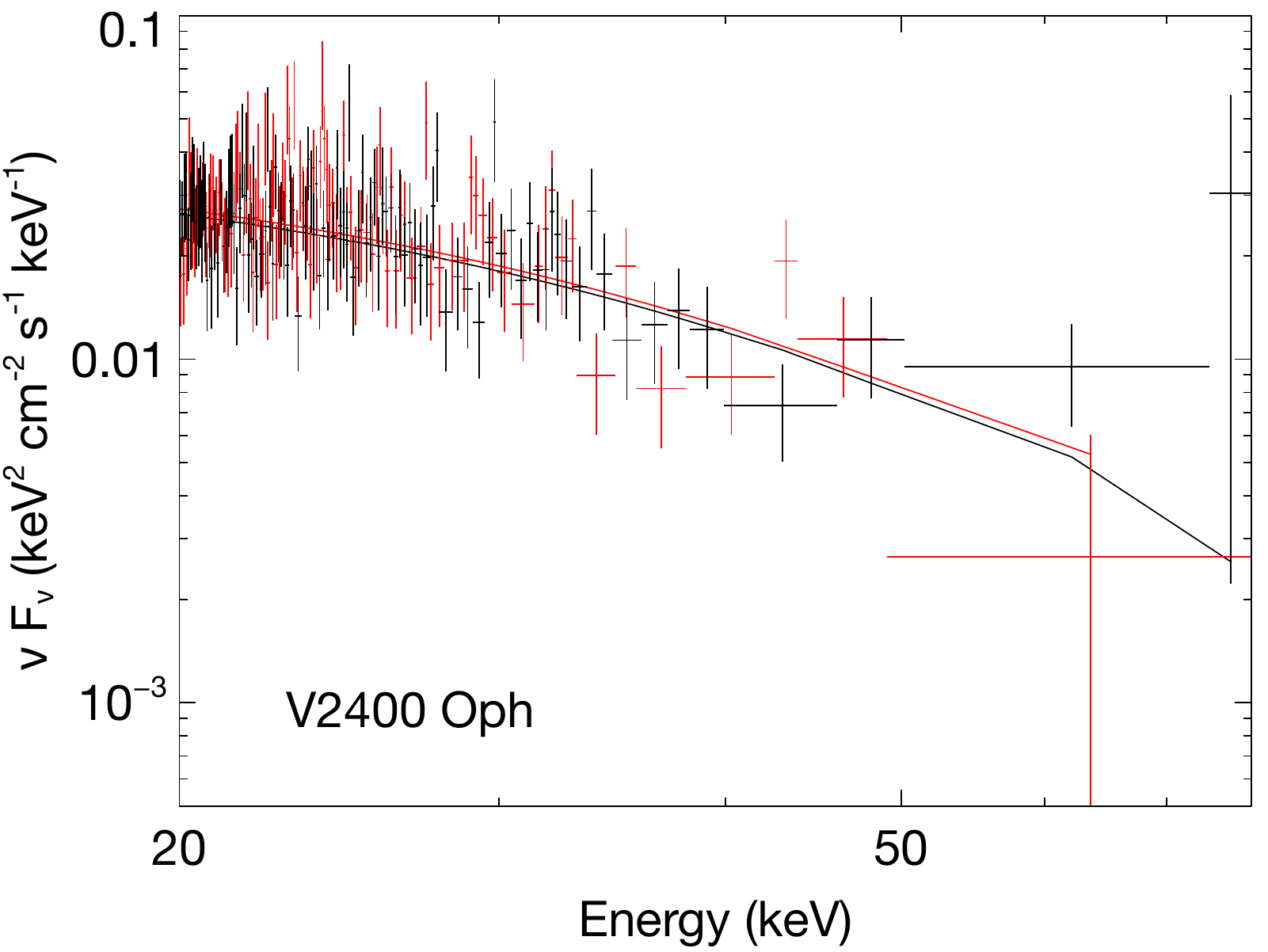}
    \end{subfigure}
    \caption{\nus\ FPMA (black) and FPMB (red) spectra of the Legacy survey targets (excluding FO\,Aqr; see Fig. \ref{fig:FO_Aqr}), fit with the \citet{Suleimanov-2019}  PSR model and plotted unfolded in $\nu$F$_{\nu}$ space. Because the IP V1062\,Tau was split into two separate observations, we fit two spectra from each FPM simultaneously. In this case we also plot FPMA (green) and FPMB (blue) of the second observation.}
    \label{fig:spec_figs}
\end{figure*}

\begin{figure*}
\ContinuedFloat
    \centering
    \begin{subfigure}[b]{0.31\textwidth}
        \includegraphics[width=\textwidth]{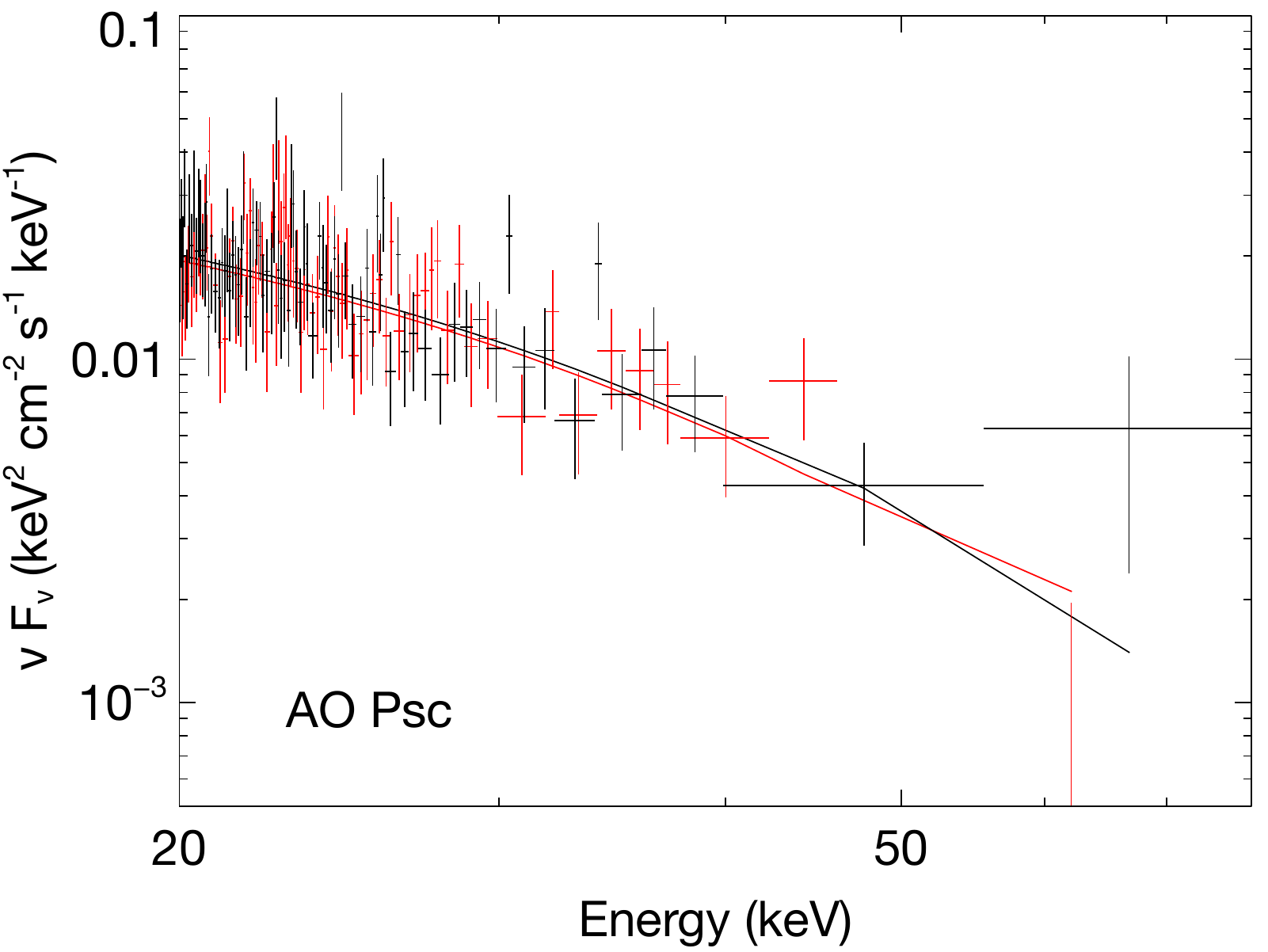}
    \end{subfigure}
    \begin{subfigure}[b]{0.31\textwidth}
        \includegraphics[width=\textwidth]{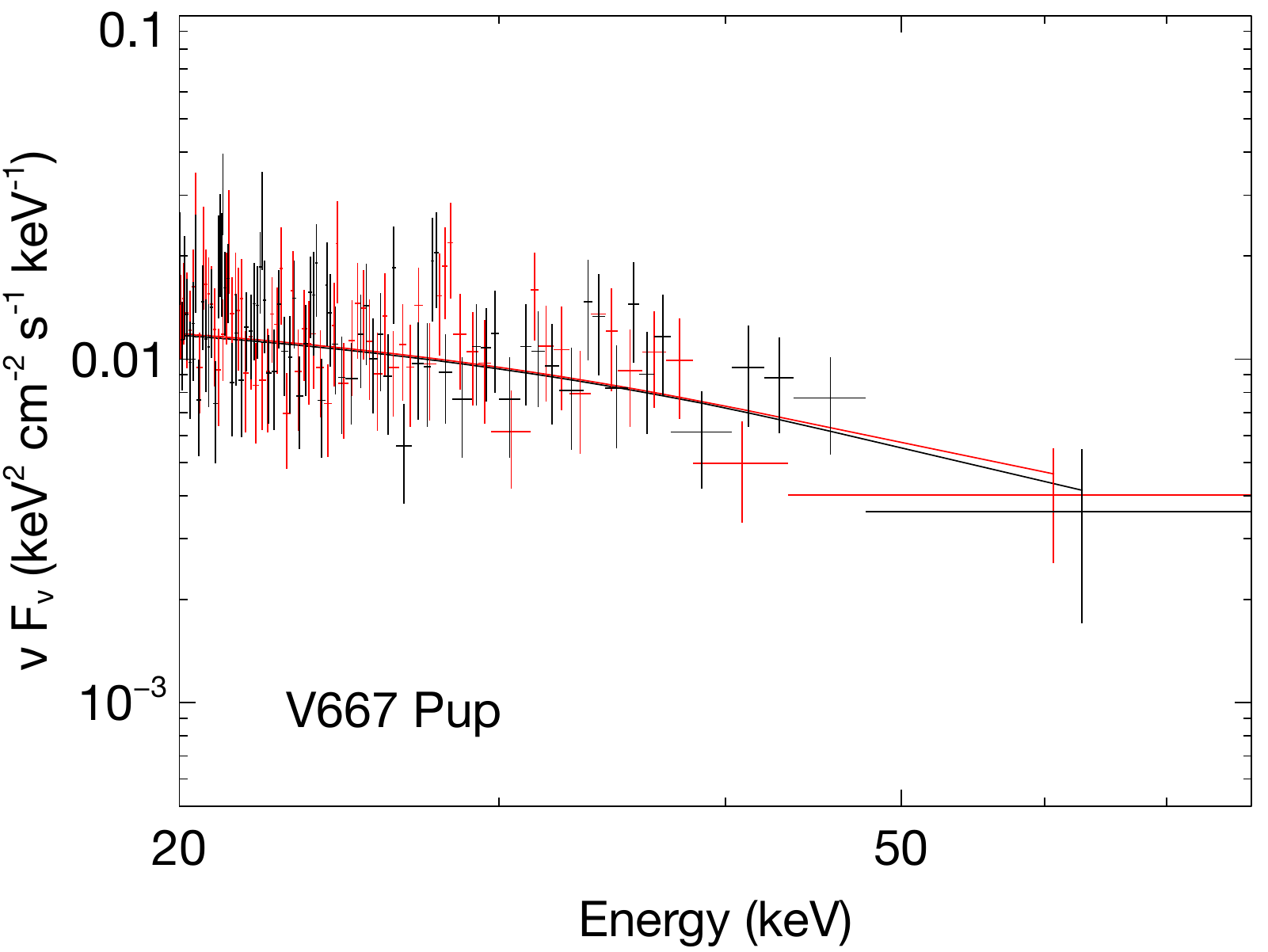}
    \end{subfigure}
    \begin{subfigure}[b]{0.31\textwidth}
        \includegraphics[width=\textwidth]{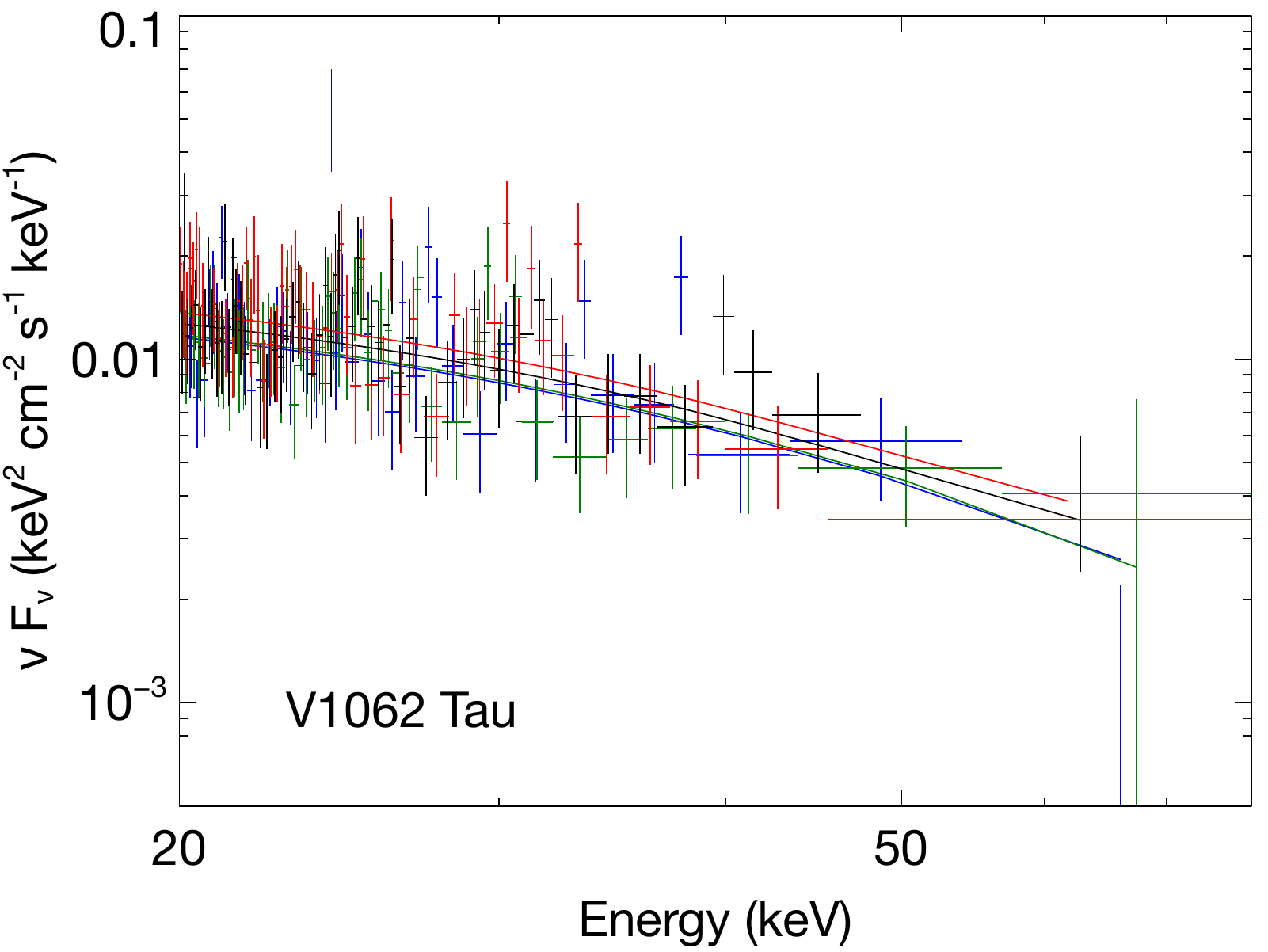}
    \end{subfigure}
    \begin{subfigure}[b]{0.31\textwidth}
        \includegraphics[width=\textwidth]{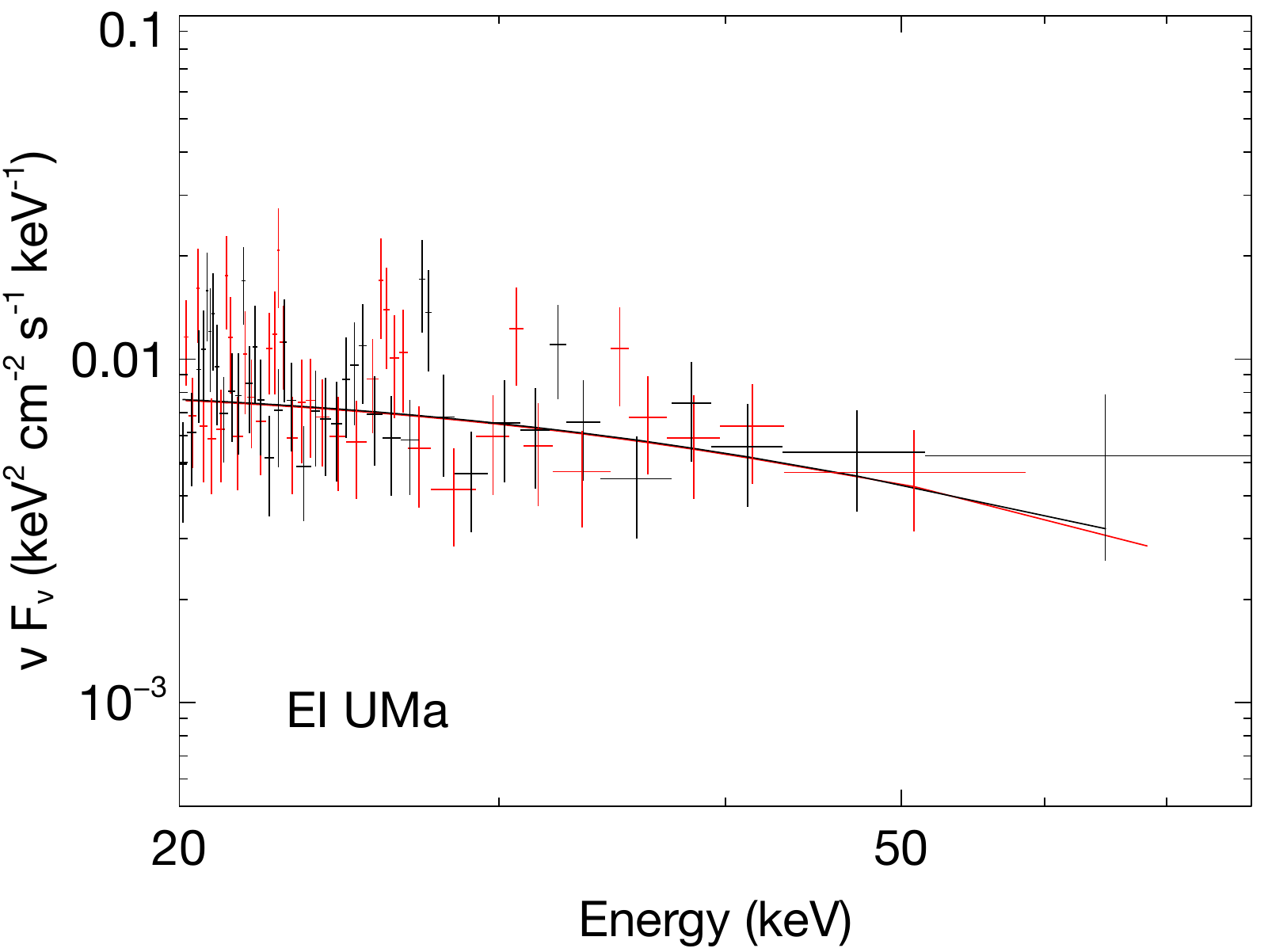}
    \end{subfigure}
    \begin{subfigure}[b]{0.31\textwidth}
        \includegraphics[width=\textwidth]{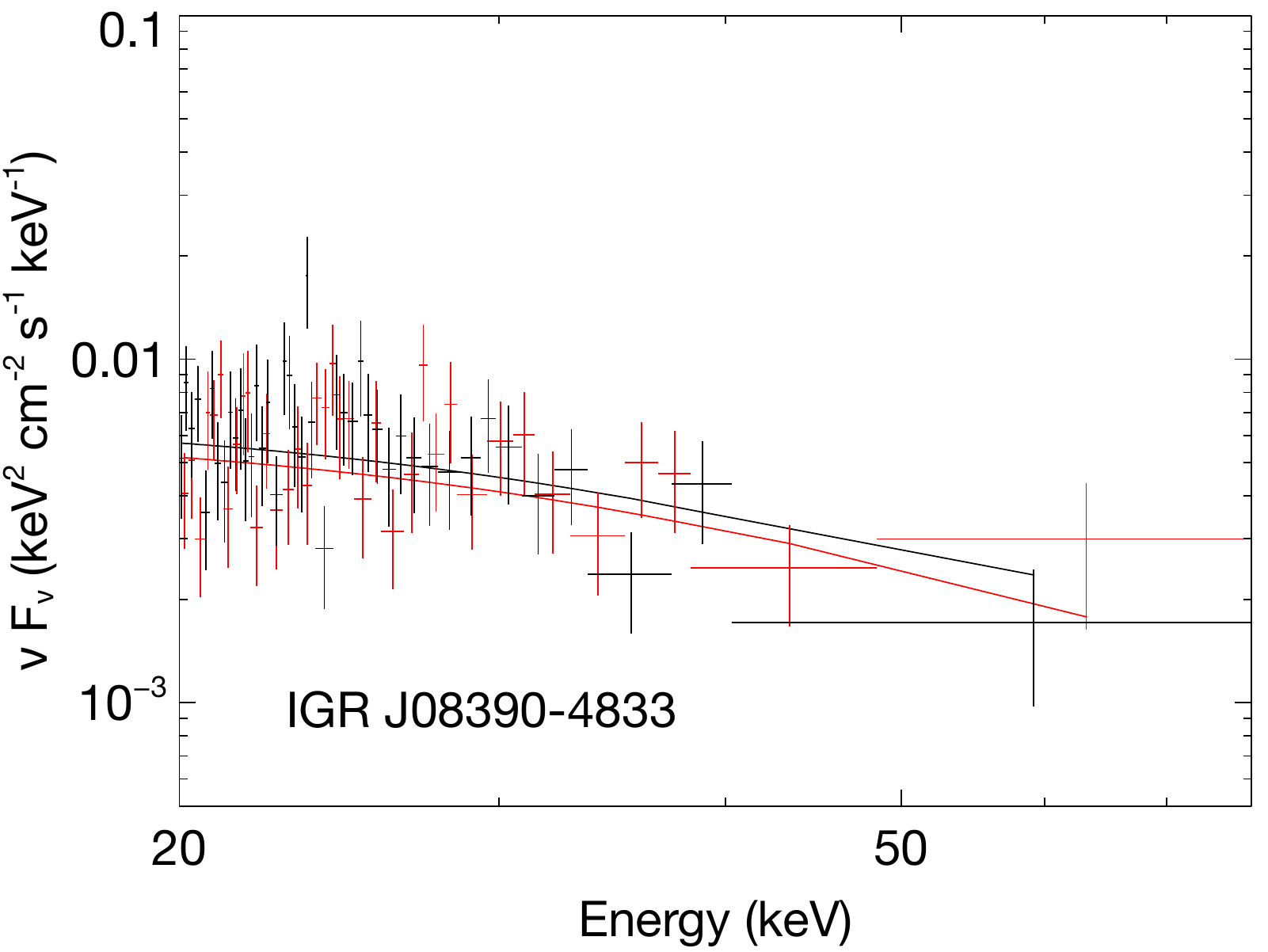}
    \end{subfigure}
    \begin{subfigure}[b]{0.31\textwidth}
        \includegraphics[width=\textwidth]{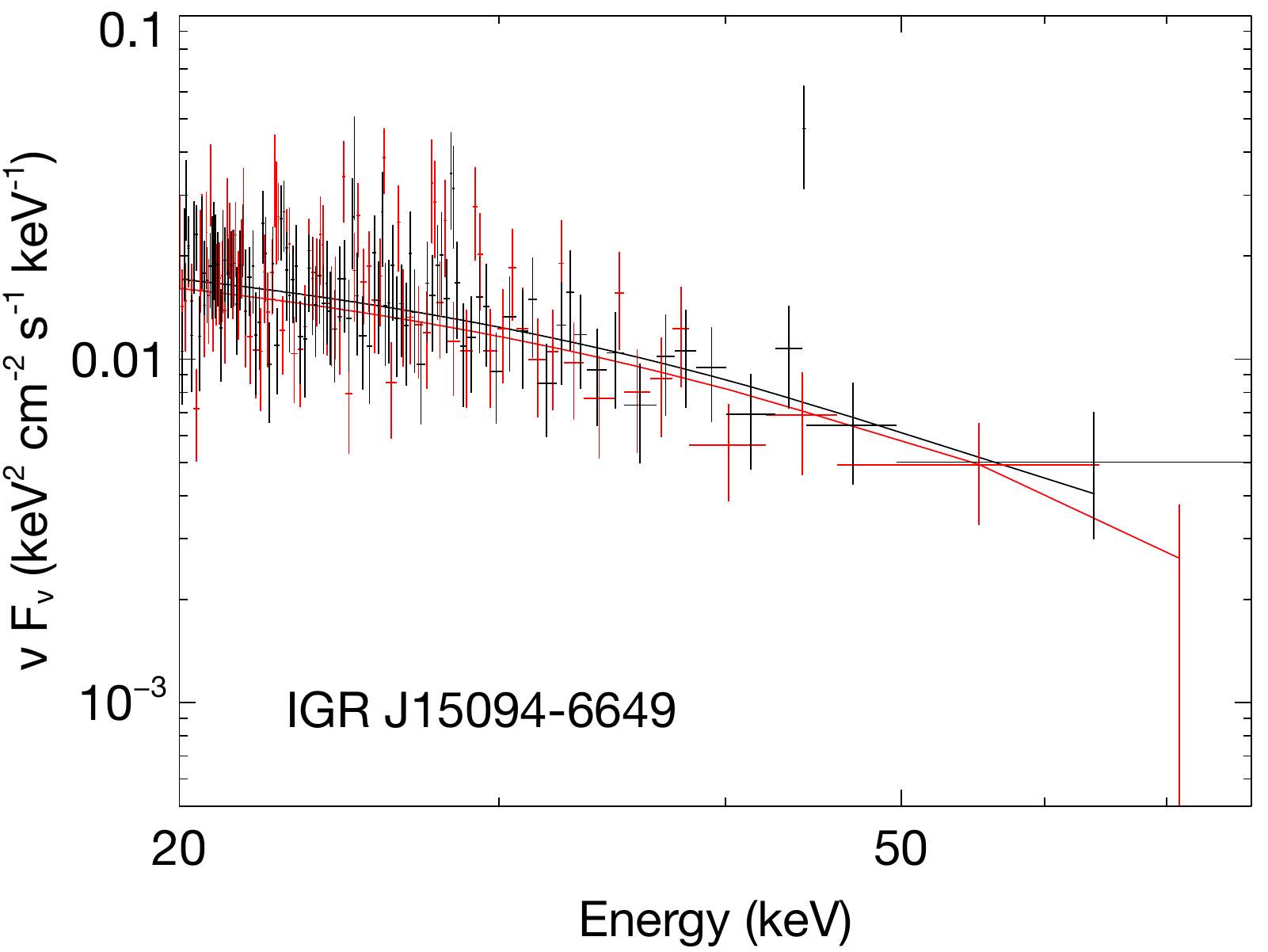}
    \end{subfigure}
    \begin{subfigure}[b]{0.31\textwidth}
        \includegraphics[width=\textwidth]{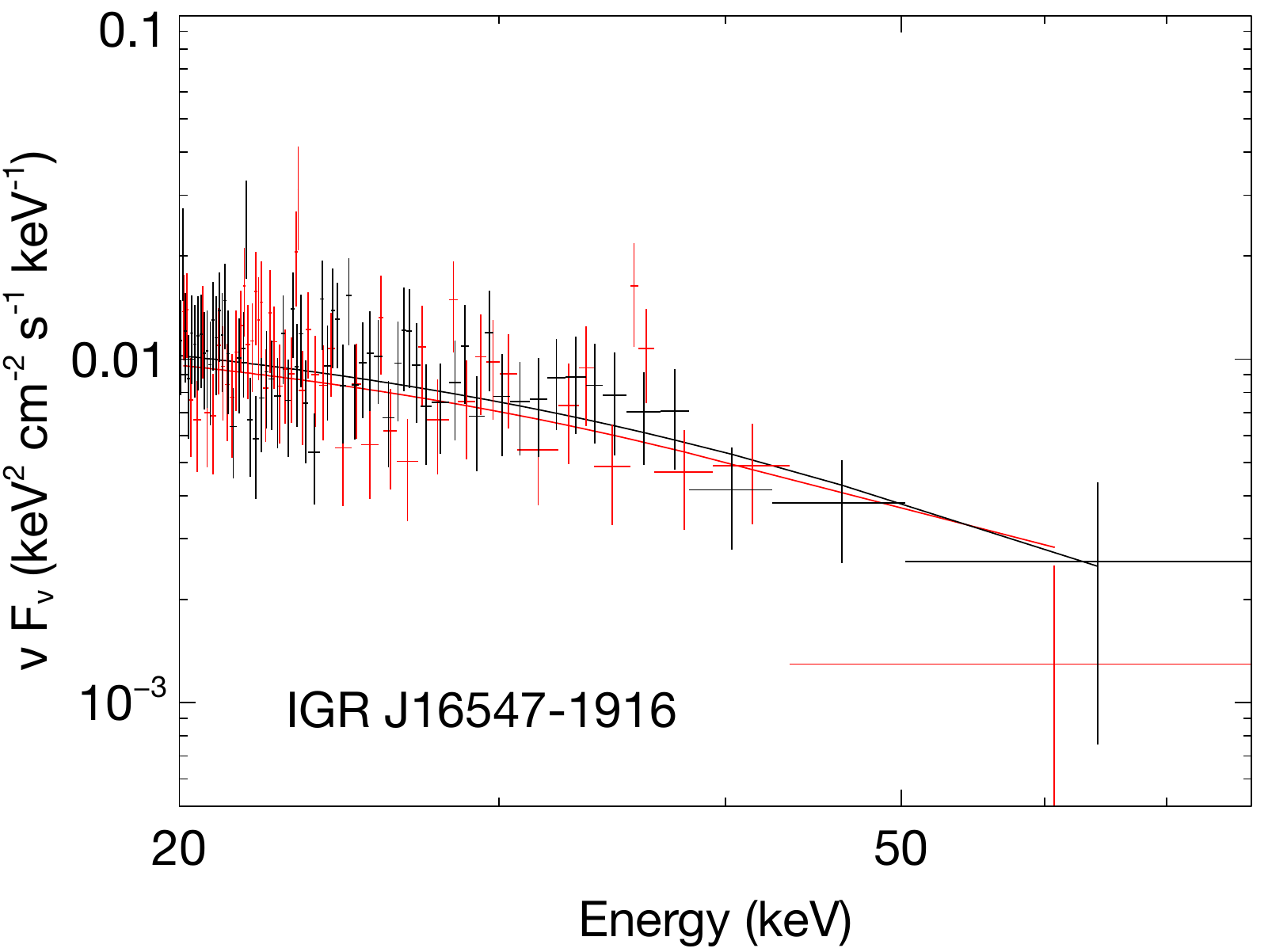}
    \end{subfigure}
    \begin{subfigure}[b]{0.31\textwidth}
        \includegraphics[width=\textwidth]{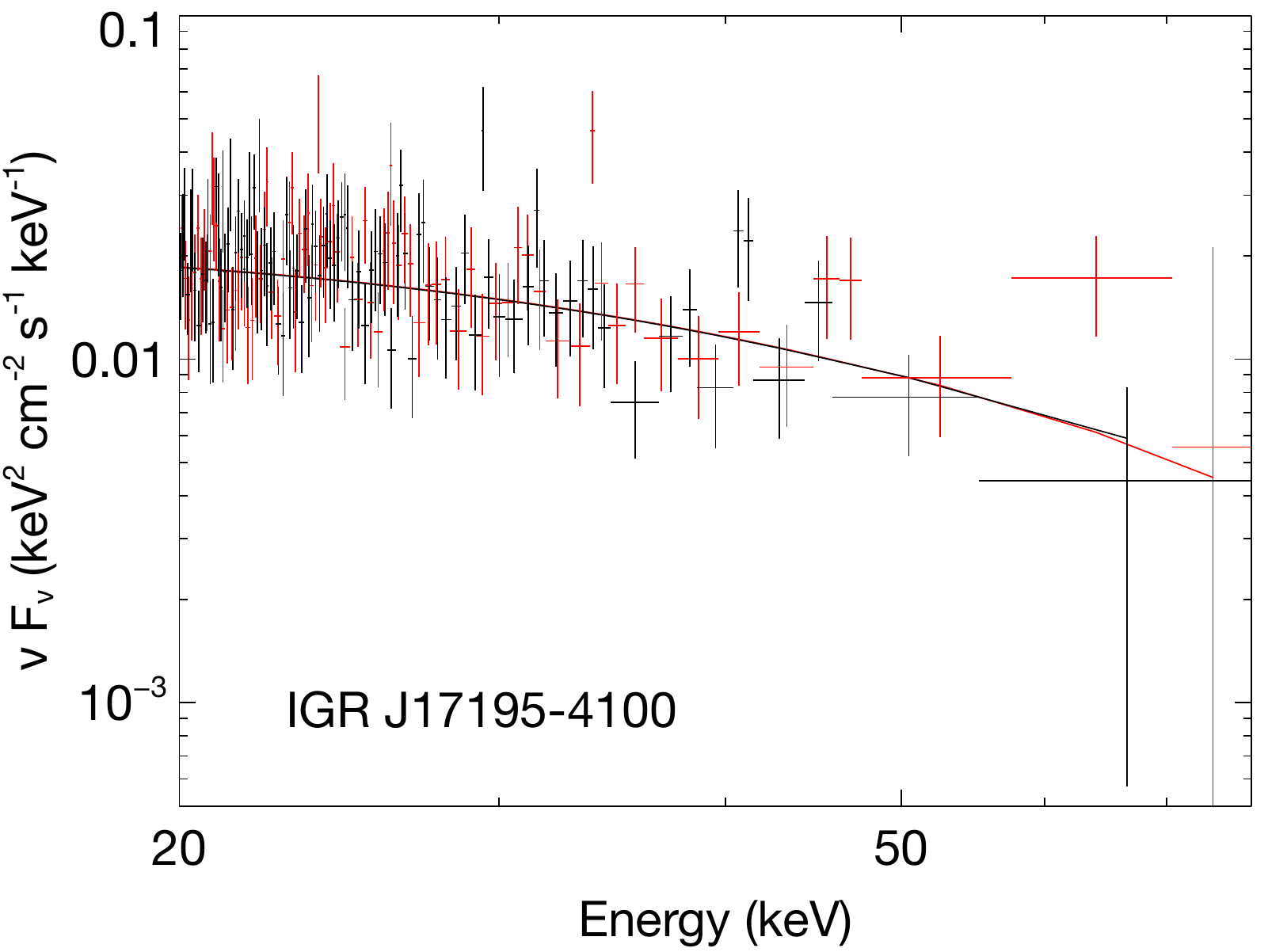}
    \end{subfigure}
        \begin{subfigure}[b]{0.31\textwidth}
        \includegraphics[width=\textwidth]{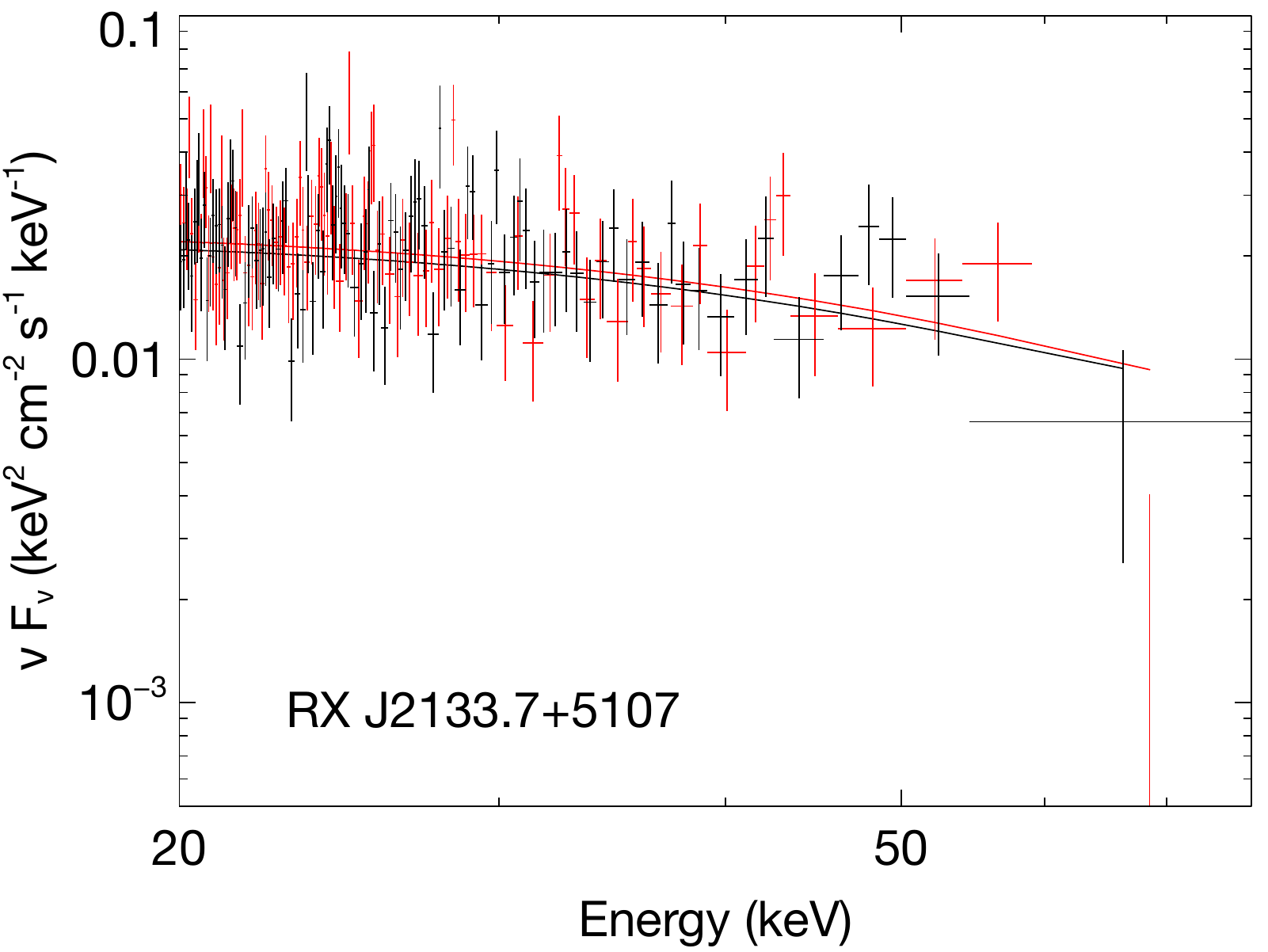}
    \end{subfigure}
    \caption{(cont.)}
\end{figure*}

% \section{Spatial distribution of the mCV sample}

% The three-dimensional distribution of the full sample of mCVs with masses constrained by \nus\ (combining the Legacy sample with that of \citealt{Suleimanov-2019}) is presented in Fig. \ref{fig:CV_map}. The distance to each object was obtained from the {\em Gaia} DR2 catalogue \citep{Bailer-Jones-2018}. The mCVs discussed in this work are isotropically distributed across the sky and range from as close as $\sim50$ pc (EX\,Hya) to as far as $\sim2.1$ kpc (IGR\,J08390$-$4833).

% \begin{figure*}
%     \centering
%     \includegraphics[width=0.9\textwidth]{Figures/mCV_distribution_mollweide.pdf}
%     \caption{The spatial distribution of all mCVs (with masses measured by \nus) discussed in this work, combining the Legacy Survey targets with the 7 previously observed with \nus\ \citep{Suleimanov-2019}. The colour of each point represents the distance as measured by Gaia DR2 \citep{Bailer-Jones-2018}. The map is in equatorial coordinates and on a Mollweide projection.}
%     \label{fig:CV_map}
% \end{figure*}

%%%%%%%%%%%%%%%%%%%%%%%%%%%%%%%%%%%%%%%%%%%%%%%%%%

% Don't change these lines
\bsp	% typesetting comment
\label{lastpage}
\end{document}